\documentclass[aps,pre,reprint,superscriptaddress,nofootinbib]{revtex4-1}
\usepackage{graphicx}
\usepackage{bm}
\usepackage{hyperref}
\usepackage[usenames,dvipsnames]{color}
\usepackage{amsmath}
\usepackage{amsfonts}

\newcommand\vb[1]{\mathbf{#1}}

\begin{document}
\frenchspacing

\title{Interplay of curvature and rigidity in shape-based models of confluent tissue}

\author{Daniel M. Sussman}\email{daniel.m.sussman@emory.edu}
\affiliation{Department of Physics, Emory University, Atlanta, GA, USA}

\date{\today}

\begin{abstract}
Rigidity transitions in simple models of confluent cells have been a powerful organizing principle in understanding the dynamics and mechanics of dense biological tissue. In this work we explore the interplay between geometry and rigidity in two-dimensional vertex models confined to the surface of a sphere. By considering shapes of cells defined by perimeters whose magnitude depends on geodesic distances and areas determined by spherical polygons, the critical shape index in such models is affected by the size of the cell relative to the radius of the sphere on which it is embedded. This implies that cells can collectively rigidify by growing the size of the sphere, i.e. by tuning the curvature of their domain. Finite-temperature studies indicate that cell motility is affected well away from the zero-temperature transition point.
\end{abstract}

\maketitle

\section{Introduction}

Recent years have seen a growing interest in the way that mechanical interactions between cells play a fundamental role in structural and dynamical processes in biology \cite{d1952growth,irvine2017mechanical}. This connection has been particularly apparent in the context of morphogenesis, where natural connections between mechanical stresses, cellular divisions, and the buckling and bending of epithelial sheets can be seen \cite{irvine2017mechanical,osterfield2013three, goodwin2019smooth,rupprecht2017geometric,streichan2018global,fouchard2020curling}. Simple coarse-grained models, ranging from lattice-based models to soft spheres to deformable polygons to phase field models \cite{Brodland2004, van2015simulating,teomy2018confluent,boromand2018jamming}, have been been useful in organizing these connections into predictive frameworks.

Here we focus on vertex models, which represent confluent monolayers as polygonal or polyhedral tilings of space; each geometrical unit corresponds to a coarse-grained cell \cite{Honda1978} and the degrees of freedom are the vertices of the geometrical units. Vertex models attempt to explicitly represent mechanical interactions between neighboring cells by force laws that depend on the local geometry of the system, and have been used to model biophysical processes covering not only morphogenesis but also wound healing and tumor metastasis  \cite{Honda2001, Bi2014,Hufnagel2006, Farhadifar2007, Friedl2009, Brugues2014, Etournay2015,spahn2013vertex}.

Such models have received attention not only for their appealing geometrical coarse-graining of clearly complex biological systems, but also for the unusual properties such models can support. For instance, two-dimensional vertex models have unusual zero-temperature rigidity transitions \cite{staple2010mechanics, Bi2015,sussman2018no,yan2019multicellular} with accompanying exotic mechanical states \cite{moshe2018geometric,noll2017active}, their glassy dynamics at finite temperature can be deeply anomalous \cite{Sussman2018epl}, and they can support unusual interfaces between coexisting populations of cells \cite{sussman2018soft}. Although systematically mapping from confluent cellular systems to these geometrical models is challenging, the models' unusual mechanical and dynamical properties suggest ways in which cells could exploit simple physical mechanisms to achieve unusual configurations or motions that may be useful for development.

\begin{figure}[b]
\centerline{
\includegraphics[width=0.45\textwidth]{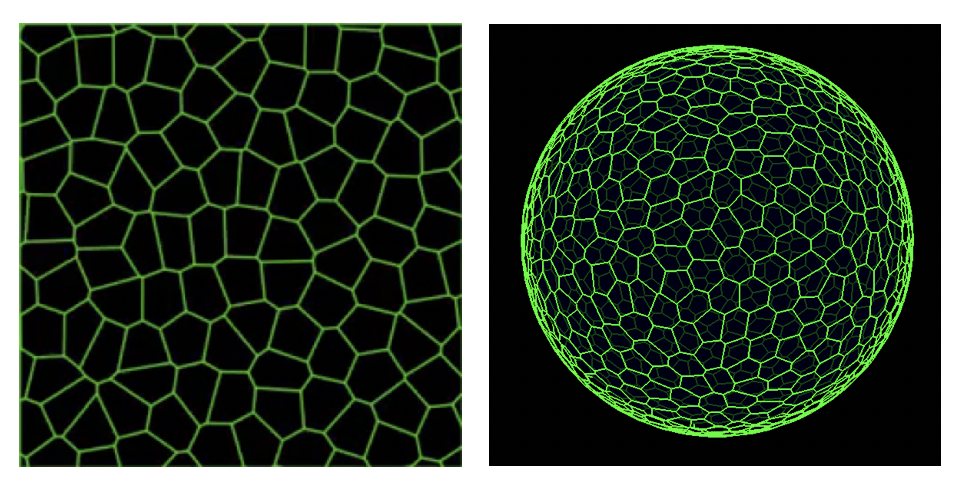}
}
\caption{\label{fig:schematic}
{\bf{Schematic image of 2D vertex models}} 
Simulation snapshots of a 2D vertex models in flat space with periodic boundary conditions (left) and embedded on the surface of a sphere (right).}
\end{figure}

While experiments on flat cellular monolayers are quite common, epithelial proliferation often takes place in domains where the \emph{curvature} of the layer is both present and may be strongly varying (as in the ellipsoidal shapes of developing embryonic systems or in the regions of both positive and negative curvature in branching morphogenesis). While gradients in curvature surely play an important role, we begin in this work by studying vertex models in domains of constant positive curvature. We are particularly interested in the interplay between the curvature of the cellular monolayer and the mechanical or dynamical state of the system. We will see that the curvature of the domain has natural consequences for the zero-temperature rigidity transition in such models, and that the finite temperature ``glassy'' behavior of cells an be strongly affected by this underlying $T=0$ transition.

There are many natural extensions of the vertex model that could be considered in moving from flat space to a three-dimensional embedding \cite{alt2017vertex,okuda2015three,Fletcher2014,osterfield2013three,murisic2015discrete}. For simplicity we consider the so-called ``3D apical vertex models'' \cite{alt2017vertex}, which represent each cell only by an apical polygon whose vertices are constrained to a surface embedded the 3D. To demonstrate the tight connection between curvature and rigidity, we focus on two-dimensional cells constrained to the surface of a sphere, as schematically illustrated in Fig. \ref{fig:schematic}, although we note that the methods described in this paper are easily extended to other holonomic constraint surfaces \cite{SAMOS}.

\section{Methods}
\subsection{An apical vertex model on the surface of a sphere}
We begin by writing  the energy functional for a flat confluent monolayer of cells,
\begin{equation}
\label{eq:energy1}
E = \sum_{i=1}^{N}{\left[k_A\left(A_i - A_0\right)^2 +k_P \left(P_i-P_0\right)^2\right]}
\end{equation}
This energy depends on the area $A_i$ and perimeter $P_i$ of each of the $N$ cells, indexed by $i$. The model parameters are the ``preferred'' geometric values, $A_0$ and $P_0$, along with the area and perimeter stiffnesses $k_A$ and $k_P$ (here we assume the monodisperse case in which all cells have identical preferences). Biologically, $A_0$ is commonly assumed to represent a combination of cellular incompressibility and the resistance of the monolayer to height fluctuations, and $P_0$ to represent a competition between tensions and adhesions acting between cells; more broadly this can be viewed as a minimal Taylor series expansion in geometrical properties that describes cellular matter rather than foams \cite{kim2018universal}. At $T=0$ a rigidity transition occurs at a particular value of $P_0$, which we denote by $P_0^*$, above which the system sits in a global energy minimum and has a vanishing shear modulus.

The density dependence of this model can be made transparent \cite{teomy2018confluent,merkel2018geometrically} by choosing the unit of length to be $\sqrt{\langle A \rangle}$ and by exploiting the fact that in these models the cells completely fill space, $\sum_i A_i = A_{total} = N\langle A \rangle$. Letting $a$ and $p$ to refer to dimensionless areas and perimeters, and letting $k_r = k_A \langle A \rangle / k_P$, Eq. \ref{eq:energy1} can be rewritten as
\begin{equation}
\label{eq:energy2}
\frac{E}{k_P\langle A \rangle } = \sum_{i=1}^{N}{\left[k_r\left(a_i - 1\right)^2 +\left(p_i-p_0\right)^2\right]} + N k_r \left(a_0-1\right)^2.\nonumber
\end{equation}
Thus, if $P_0$ is a control parameter (\emph{and} if the stiffnesses $k_r$ and $P_0$ are themselves density-independent, an interesting biological question), the density dependence of the model in flat space enters via $p_0 = P_0/\sqrt{\langle A \rangle} = P_0 \rho^{1/2}$. Writing this dimensionless form for the energy makes it clear that the parameter $a_0$ couples to the total size of the system -- serving as an offset to the total pressure -- but it does not affect the forces between degrees of freedom \cite{teomy2018confluent,merkel2018geometrically} and hence does not control the rigidity transition. To be explicit, via this mechanism if cells divide in a domain of fixed area their average size decreases and cells which were initially in an \emph{incompatible} regime of parameter space ($p_0 < p_0^*$, in which geometrical frustration prevents the cells from achieving the global energy ground state) could enter into a \emph{compatible} regime of parameter space. Thus, in flat space cells in this model can unjam via growth.

Extending the above expressions to a \emph{spherical vertex model} requires no change of notation (although justifying the geometric coarse-graining would require a more biologically informed derivation, as we discuss in the conclusion). We simply interpret the ``areas'' and ``perimeters'' to be those measured on the sphere: perimeters are given by sums of geodesic distances as one traverses the vertices composing the cell, and areas are given by the area of the spherical polygons enclosed by those geodesic arcs. The forces acting on the vertices are given by the negative spherical gradients of Eq. \ref{eq:energy1} (explicit expressions are given below). The statistical mechanics of fluids confined to curved manifolds is itself a rich topic \cite{tarjus2012statistical,guerra2018freezing}, and a natural non-biological application of the methods developed here are to general phenomena of disordered rigidity transitions in non-Euclidean spaces.

To implement efficient and highly scalable numerical simulations of the above equations, allowing T1 transitions to facilitate neighbor exchanges between cells and evolving the degrees of freedom under equations of motion ranging from energy minimization schemes to overdamped Brownian dynamics to self-propelled ``active'' dynamics, we combine the GPU-accelerated frameworks described in Refs. \cite{sussman2017cellGPU,sussman2019fast}. We now provide some additional computational details for the interested reader.

\subsection{Projection operator formalism for the constraint surface}
We follow Refs. \cite{SAMOS,sknepnek2015active} in using the projection operator formalism to enforce the hard constraint that the degrees of freedom lie on the surface of a sphere. Explicitly, for the case of overdamped Brownian dynamics at temperature $T$ we write
\begin{equation}
\Delta \vb{r}_i = P_T\left( \vb{r}_i, -\mu \Delta t \nabla_i E + \eta_i \right),
\end{equation}
where $\mu$ is an inverse friction coefficient, $\eta$ a normally distributed random force with zero mean and with $\langle \eta_{i\alpha}(t) \eta_{j\beta}(t') \rangle = 2\mu T\Delta t \delta_{ij}\delta_{\alpha\beta}$ in each of the three Cartesian directions denoted by greek indices (so that the noise has the correct statistics in the tangent plane of the vertex). The operator $P_T(\vb{a}, \vb{b}) = \vb{b} - \left(\hat{\vb{a}}\cdot\vb{b}\hat{\vb{a}}\right)$ projects the forces and the random noise onto the tangent plane at the location of the degree of freedom. To maintain the spherical constraint small time steps must be used, and degrees of freedom are projected back onto the surface of the sphere of radius $R$ after they are moved via $\vb{r}_i(t+\Delta t) = P_N(\vb{r}_i(t) + \Delta \vb{r}_i)$ for $P_N(\vb{a}) = R \frac{\vb{a}}{|\vb{a}|}$.

\subsection{Explicit force calculations}
For completeness, and to better illustrate the computational challenges that must be addressed when simulating 2D vertex models constrained to curved surfaces, we explicitly document some of the expressions used to compute forces in the vertex model where the degrees of freedom are constrained to lie on the surface of a sphere of radius $R$. As in previous works \cite{Bi2015,sussman2017cellGPU,bi2016motility}, the calculation of the gradient of  Eq. \ref{eq:energy1} is readily expanded via chain rules to to separate out contribution from the particular functional form of the energy and from the entirely geometric quantities, i.e., how much distances and areas of polygons change when degrees of freedom are moved. Thus, the new quantities to implement in the present case of a spherical vertex model are a complete set of primitives for calculating geodesic distances, spherical polygon areas, and the appropriate derivatives of each with respect to vertex positions.

We note that on the surface of the sphere there are many equivalent expressions for the geodesic distance between two points (or the included angle between three points, or the area of a spherical polygon given by $n$ points, etc). While \emph{analytically} equivalent, these expressions typically have different regimes of \emph{numerical} stability. For instance, given two points on the sphere, $\vec{n}_1$ and $\vec{n}_2$, the distance $d$ may be written as
\begin{equation}
d(\vec{n}_1,\vec{n}_2) = \left\{
	\begin{aligned}
	d_a&=R \cos^{-1}\left(\hat{n}_1\cdot\hat{n}_2\right)\\
	d_b&=R \sin^{-1}\left(\left|\hat{n}_1\times \hat{n}_2\right|\right)\\
	d_c&=R \tan^{-1}\left(\frac{\left|\hat{n}_1\times \hat{n}_2\right|}{\hat{n}_1\cdot\hat{n}_2}\right)\\
	\end{aligned}
\right. .
\end{equation}
The first expression above is the simplest and least computationally expensive, but it is poorly conditioned for very small distances (as might be relevant when vertices get very close to each other before performing a T1 transition). The second expression is poorly conditioned for large distances, whereas the third is the most computationally expensive but is well-conditioned for all distances. These questions of numerical stability become especially acute when dealing with the forces, and we have found it important to implement self-consistency checks on the force calculations and switch to analytically equivalent but numerically different routes of calculating gradients in the spherical vertex model.

\subsubsection{Gradient calculations}
Given a vertex position $\vec{n}_1$, which our program stores in $\mathbb{R}^3$, we first express it in the usual spherical basis $\vec{n}_1 = \{r_1,\theta_1,\phi_1\}$ and compute the local $\hat{\theta}$ and $\hat{\phi}$ directions. The spherical gradient of the distance between two vertices as the position of the first vertex is changed is then given by
\begin{equation}
\nabla_1 d(\vec{n}_1,\vec{n}_2) = \frac{1}{R} \frac{\partial d}{\partial \theta_1}\hat{\mathbf{\theta}}_1 +\frac{1}{R\sin\theta_1} \frac{\partial d}{\partial \phi_1}\hat{\mathbf{\phi}}_1,
\end{equation}
where choosing a particular formula to compute the geodesic distance we have
\begin{align}
d_a(\vec{n}_1,\vec{n}_2) = R \cos^{-1}&\big( \cos(\theta_1)\cos(\theta_2)  \\&+ \sin(\theta_1) \sin(\theta_2)\cos(\phi_1-\phi_2)\big). 
\end{align}
Thus,
\begin{widetext}
\begin{eqnarray}
\frac{1}{R} \frac{\partial d_a}{\partial \theta_1} &=& \frac{\cos(\theta_2)\sin(\theta_1) - \cos(\theta_1) \cos(\phi_1-\phi_2) \sin(\theta_2)}{\sqrt{1-\left(\cos(\theta_1)\cos(\theta_2) + \cos(\phi_1-\phi_2) \sin(\theta_1)\sin(\theta_2) \right)^2}}, \\
\frac{1}{R\sin\theta_1} \frac{\partial d_a}{\partial \phi_1}&=& \frac{\sin(\theta_2) \sin(\phi_1-\phi_2)}{\sqrt{1-\left(\cos(\theta_1)\cos(\theta_2) + \cos(\phi_1-\phi_2) \sin(\theta_1)\sin(\theta_2) \right)^2}}.
\end{eqnarray}
\end{widetext}
From this one readily appreciates the substantial cost of computing gradients of $d_b$ or $d_c$, and so whenever possible we opt for the simpler expressions stemming from $d_a$.

Similarly, that the area of a spherical triangle $A(\vec{n}_1,\vec{n}_2,\vec{n}_3)$ can be written as
\begin{eqnarray}
 A&=&R^2\left(\alpha+\beta+\gamma -\pi \right), \textrm{ where}\\
\alpha &=& \cos^{-1} \left( \frac{\cos(a) - \cos(b)\cos(c)}{\sin(b)\sin(c)}\right), \\ 
\beta &=&\cos^{-1} \left( \frac{\cos(b) - \cos(a)\cos(c)}{\sin(a)\sin(c)}\right),\\
\gamma &=&\cos^{-1} \left( \frac{\cos(c) - \cos(a)\cos(b)}{\sin(a)\sin(b)}\right),\\
\end{eqnarray}
where 
\begin{eqnarray}
a &=& d(\vec{n}_2,\vec{n}_3)/R,\\
b &=& d(\vec{n}_1,\vec{n}_3)/R, \\
c &=& d(\vec{n}_1,\vec{n}_2)/R.
\end{eqnarray}
Clearly, again, care must be taken in choosing distance functions that will lead to well-conditioned expressions for both the area and gradients of the area while also minimizing the complexity of the resulting expressions. Additional considerations include the efficiency and numerical stability of computing cellular areas as either the sum of spherical triangles formed by the cell centroid and consecutive vertices around the cell or via the sum of the included angle at each of the $n$ vertices,
\begin{equation}
A(\{\vec{n}_1, \vec{n}_1,,\ldots,\vec{n}_n,\}) = R^2\left( \left(\sum_i^n \alpha_i\right) -(n-2)\pi\right);
\end{equation}
this is particularly delicate when the cells are not convex, or when edges cross.

\subsection{Initial conditions}
The initial conditions for both the zero-temperature quenches and the finite-temperature simulations reported in this work are chosen to be high-temperature random configurations of cells. We create these configurations by first picking a desired number of cells, $N_c$, and distributing $N_c$ points uniformly on the surface of the sphere. We use the Computational Geometry Algorithms Library (CGAL) \cite{cgal:eb-19b,cgal:hs-ch3-19b} to construct the convex hull of these points, and take the initial vertex positions to be the centroids of the resulting facets (projected back onto the sphere) \cite{fortune1995voronoi}.

\section{Results}

\subsection{Athermal rigidity transition}
We first directly probe the athermal rigidity transition of the spherical vertex model as a function of $N$ and $p_0$; for simplicity here we first focus on the $k_r=0$ limit of Eq. \ref{eq:energy2}. We prepare between 100 and 500 initial configurations for each value of $p_0$, seeded by randomly placing cell centers on the surface of the sphere and deriving the initial positions of the vertices from the convex hull of that point pattern. We perform a FIRE  energy minimization \cite{bitzek2006structural} of these configurations  to find the inherent state associated with each initial configuration. Like its counterpart in flat space, the spherical vertex model described here is extensively underconstrained; as such, we anticipate that the ground states of the model are mechanically stable only in the presence of residual stresses \cite{moshe2018geometric,sussman2018no,merkel2018geometrically}. 

\begin{figure*}[bht!]
\centerline{
\includegraphics[width=0.5\textwidth]{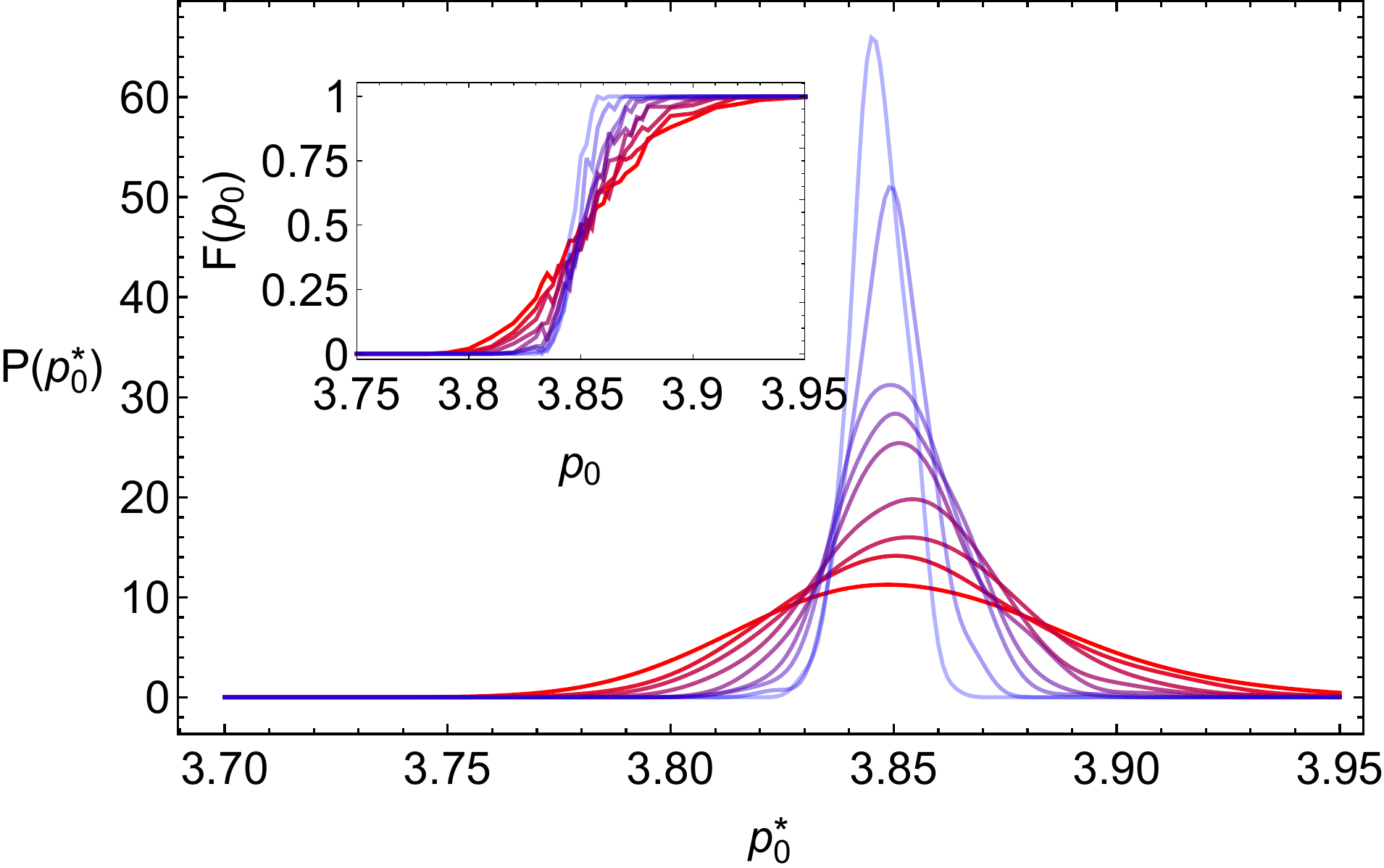}\includegraphics[width=0.5\textwidth]{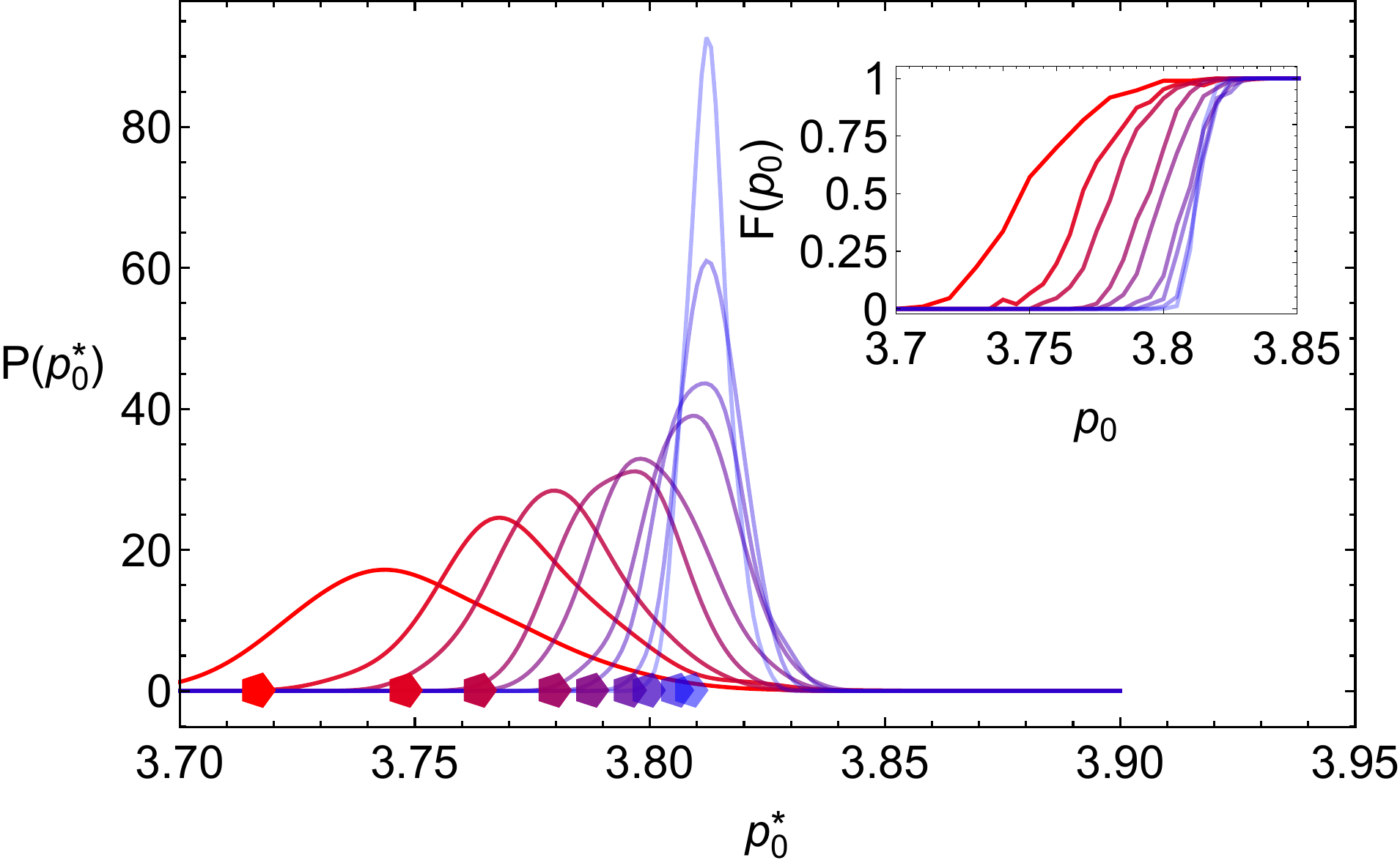}
}
\caption{\label{fig:transitionDistribution}
{\bf{Finite size effects in vertex model transitions for flat monolayer (left) and spherical (right) monolayers}} 
Probability distribution of the critical value of perimeter, $p_0^*$, above which the $k_r=0$ vertex model transitions from a rigid to a mechanically unstable system. The main plots show the derivative of the fraction of states at zero energy, $F(p_0)$ (shown in the insets). Colors correspond to system sizes $N=32,\ 48,\ 64,\ 96,\ 128,\ 192,\ 256,\ 512,\ 1024$ (darker red to lighter blue). The correspondingly colored pentagons show the perimeter of a unit-area regular pentagon on a sphere of radius  $\sqrt{N/(4\pi)}$.}
\end{figure*}

Thus, we estimate the rigidity transition for a given value of $N$ by computing the fraction of minimized states, $F(p_0,N)$, which minimize to an inherent state of zero energy. The probability distribution of transition points is given by the derivative of this function; to take this derivative while suppressing noise, we convolve a linear interpolation of the $F(p_0,N)$ with the derivative of a Gaussian whose standard deviation is related to the shape of $F(p_0,N)$ (see Ref. \cite{merkel2018geometrically}). We have done this for both for the planar and spherical vertex models at $k_r=0$, and the results are shown in Fig. \ref{fig:transitionDistribution}.

Our results for the mean value of the transition for the planar vertex model, and the variance of the distribution, are consistent with previous studies \cite{merkel2019minimal} (although note that other simulations, based on the Surface Evolver package \cite{brakke1992surface} and minimizing under a different protocol, have reported slightly different results \cite{Bi2015}). As might be expected, the primary effect of approaching the thermodynamic limit in the planar case is to develop a more sharply peaked distribution about the $N\rightarrow\infty$ limiting value, with very little change in the mean value of the distribution. In contrast, the effects of changing the size of the sphere relative to the typical size of each cell is readily seen in the way the distribution of the transition point not only sharpens but also shifts with $N$.

The critical value of $p_0$ separating the mechanically rigid and floppy phases as a function of $N$ closely tracks (but is not precisely equal to) the way in the which the perimeter of a unit-area regular pentagon varies on a sphere of total surface area $N$. This value of $p_0$ forms a natural bound for the \emph{non-linear} rigidity transition: sufficiently large cellular displacements require cells to exchange neighbors, on average cells have six sides, so during the T1 transition a spherical pentagon must be formed. If $p_0(N) < p_{penta}(N)$ this configuration will cost energy, but the precise connection between this bound on the nonlinear behavior of rearrangements and the infinitesimal rigidity calculation shown in Fig. \ref{fig:transitionDistribution} remains unclear, both here and in the planar case \cite{moshe2018geometric}.

\subsection{Finite temperature dynamics}
To show that this qualitative shift is neither just a result of the $k_r=0$ limit explored above nor an artifact of the exotic mechanical states at zero temperature found in vertex models (i.e., a result just of studying an extensively underconstrained model) \cite{noll2017active, sussman2018no,merkel2019minimal,moshe2018geometric}, we study the finite-temperature dynamics of disordered configurations of the spherical vertex model, with a range of $k_r$, $p_0$, $N$, and $T$. We have performed overdamped Brownian dynamics, and to illustrate the importance of curvature here we present representative data in which we consider fixing $p_0 = 3.775$ (less than the planar critical value, but above the spherical critical value for small $N$) while varying both $N$ and $T$ (while holding the typical cell size fixed, so that varying $N$ corresponds to varying the curvature of cellular substrate). 

The characteristic relaxation time of the simulated systems, $\tau_\alpha(T,p_0)$,  were calculated as in Ref. \cite{Sussman2018epl} using the decay of the self-overlap function \cite{keys2007measurement}. This function measures the fraction of particles that have been displaced by more than a characteristic distance $b$ after a time $t$,
\begin{equation}
Q_s(t) = \frac{1}{N}\sum_{i=1}^N w \left( |\vec{r}_i(t) - \vec{r}_i(0) | \right),
\end{equation}
where $\vec{r}_i$ is the vector position of cell $i$, $w$ is a window function, $w(r \leq b) = 1$ and $w(r>b) = 0$. The cutoff $b$ plays a very similar role to a choice of  $\vec{q}$ when looking at the decay of the self-intermediate scattering function,
\begin{equation}
F_s(q,t)=N^{-1} \left\langle \sum_i e^{ i \vec{q}\cdot \left(\vec{r}_i(t) - \vec{r}_i(0)\right) } \right\rangle.
\end{equation}
We choose $b = 1/2$, and estimate $\tau_\alpha(T,p_0)$ as the time it takes for $Q_s(t)$ to decay to $1/e$.

\begin{figure}[b]
\centerline{
\includegraphics[width=0.45\textwidth]{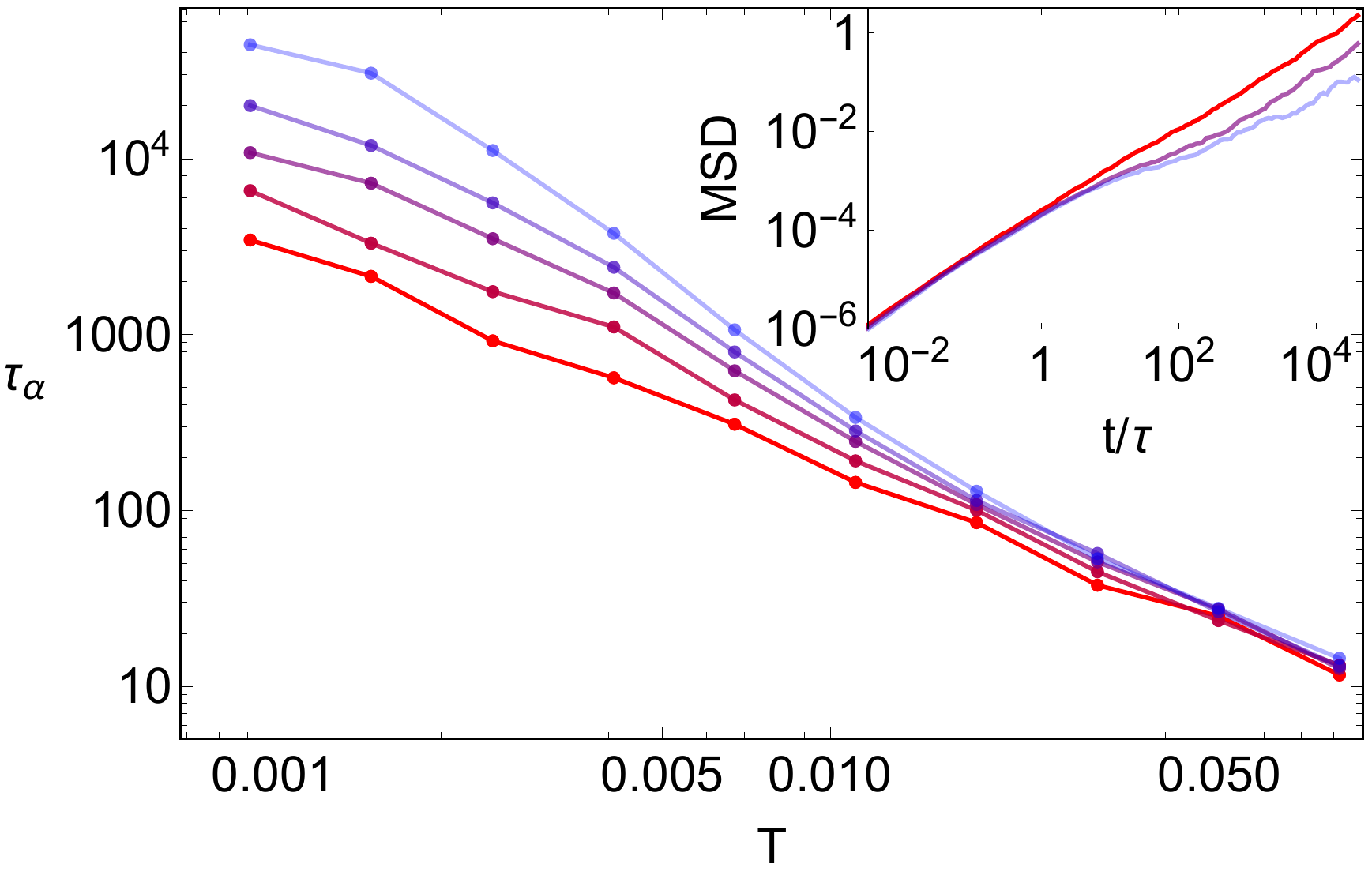}
}
\caption{\label{fig:msd}
{\bf{System size takes one from a fluid to a glassy regime}} The $\alpha$-relaxation time as as function of temperature for $N=32,\ 128,\ 1024$ (dark red to light blue), $p_0=3.775$, and $k_r=0$ goes from fluid-like to glassy as $N$ increases. Data averaged over 2-10 independent simulations for larger and smaller $N$. (Inset)  The mean-squared displacement, in units of $\langle A \rangle$, in the spherical vertex model with $p_0=3.75$, $k_r=1$, and $N=32,\ 128,\ 1024$ (dark red to light blue) averaged over $(8192/N)$ independent simulations at $T=5\times 10^{-4}$. Again, the MSD shows a transition from purely diffusive for small $N$ to transiently caged for large $N$.}
\end{figure}

Figure \ref{fig:msd} shows representative changes in how the $\alpha$ relaxation time grows both with decreasing temperature and with increasing $N$. One clearly sees how the underlying change in $p_0^*(N)$ at $T=0$ affects the finite-temperature dynamics, with these examples showing an order-of-magnitude change in the dynamical scale (measured either as a time scale or a magnitude of typical displacements) as the system size changes from $N=32$ to $N=1024$. Notably both the MSD and relaxation-time data indicate a qualitative change in the temperature dependence of the dynamics as the system size is increased: the MSD data in the inset shows an example of cellular motions crossing over from simple diffusive behavior for small $N$ to caged, glassy behavior as $N$ is increased, and this is reflected in the curvature visible in the plot of  $\log \tau_\alpha(T)$ for large $N$ but not for small $N$. We note that the \emph{magnitude} of these dynamical effects will depend on $\left(p_0-p_0^*(N\rightarrow\infty)\right)$; a systematic study of these effects is currently underway.

\section{Discussion}

The fact that the mean of the rigidity transition shifts as the relationship between curvature and cell-size varies suggests a novel mechanism by which cells could collectively tune between different mechanical phases as a function of their curvature. Models of 3D collections of cells in embryonic zebra fish development have shown the potential for coexistence between fluid-like behavior in regions of high curvature and solid-like behavior in regions of lower curvature \cite{mongera2018fluid}. Perhaps more relevant to this explicitly two-dimensional model, developing insect embryos look much like ellipsoidal versions of the right panel in Fig. \ref{fig:schematic}, with regions of high and low curvature. Thus, although we currently neglect gradients in curvature, the curvature-dependent rigidity discussed here might be directly relevant in the modeling of such systems \cite{jain2019regionalized}.

This connection between curvature and the ability to support mechanical stresses suggests a relationship between density and jamming that is \emph{qualitatively different} from the planar case. There is now a competition between the scaling of the critical perimeter with typical cell size and the effect of curvature as expressed by ratio of the sphere radius to the typical cell size, leading to $P_0^*(N) = \langle A\rangle^{1/2} p_0^*(N)$, where $p_0^*(N)$ itself is estimated from the mean of the distributions in Fig. \ref{fig:transitionDistribution}.

\begin{figure}
\centerline{
\includegraphics[width=0.5\textwidth]{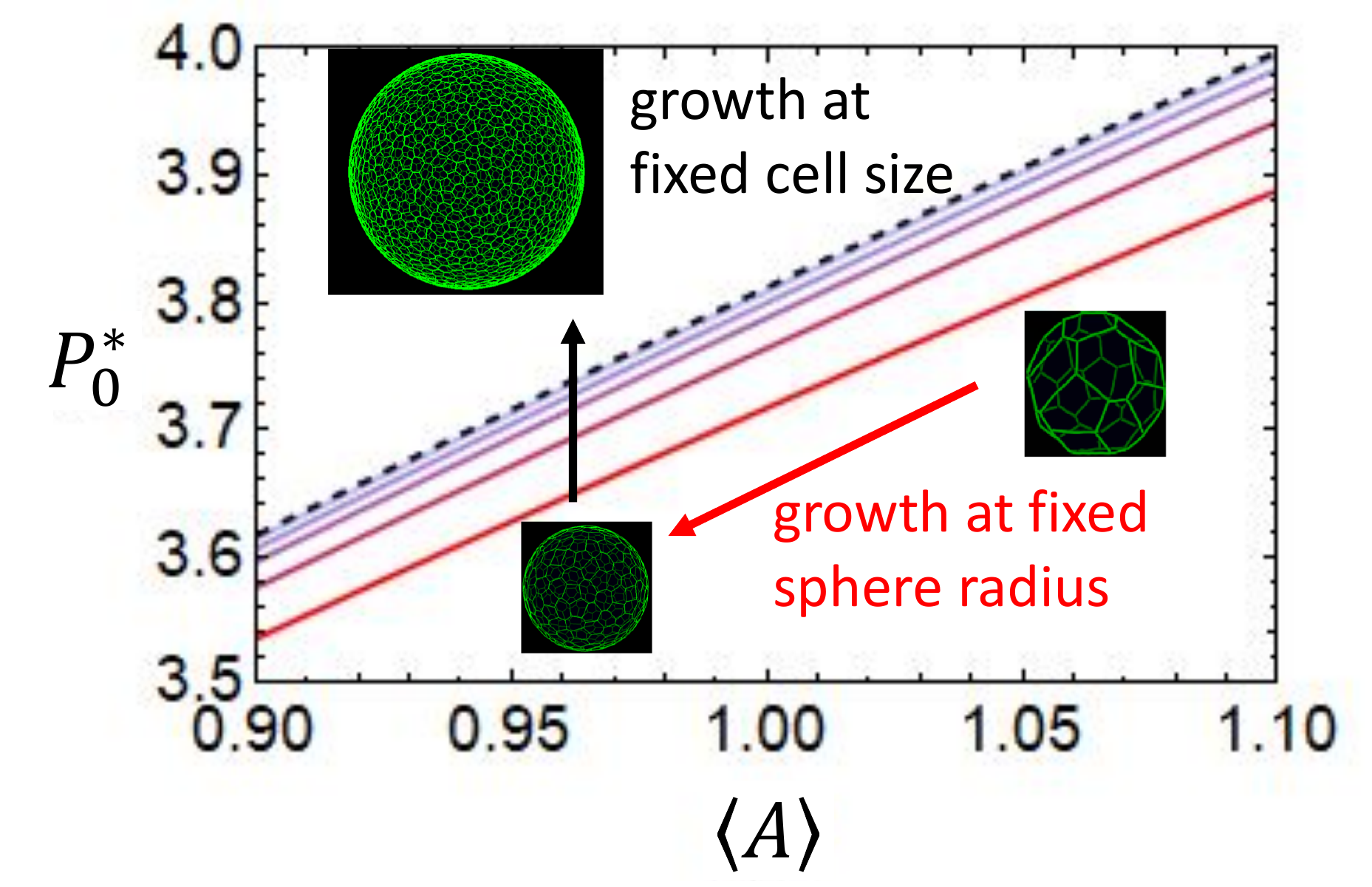}
}
\caption{\label{fig:criticalPerimeter}
{\bf{Jamming and unjamming via cellular division and death}} 
Curves show the estimated dependence of $P_0^*$ as a function of $\langle A \rangle$ in the spherical vertex model for $N=32,\ 64,\ 128,\ 256$ (dark red to light blue) and for $N=1024$ (dashed black line) all at $\langle A \rangle = 1$. The value $p_0^*(N)$ is estimated from the mean of the distributions in Fig. \ref{fig:transitionDistribution}, on top of which the energy functional dependence of $\sqrt{\langle A \rangle}$ is applied. Schematically, cells dividing in a finite fixed domain would correspond to decreasing the typical cell size, moving the system along curves of constant color, whereas cells dividing in a growing domain would move the system across curves of different colors. This shows the possibility of cells collectively \emph{unjamming} or \emph{jamming} via growth at fixed other model parameters.}
\end{figure}

Crucially, the effect of curvature is strong enough to \emph{reverse} the qualitative dependence on density. This is schematically depicted in Fig. \ref{fig:criticalPerimeter}. which shows an estimate of the shifting of the rigidity transition $P_0^*(N,\langle A\rangle)$. In one limiting case, the number of cells could increase on a sphere of fixed radius. In this scenario, the decrease of $\langle A \rangle$ lowers the critical transition point, so cells dividing (at constant other model parameters) could induce the system to collectively \emph{unjam via growth}. In the other limiting case, the number of cells could increase in a simultaneously enlarging spherical domain (so that the cell number increases at fixed $\langle A \rangle$). In such a case the system could potentially \emph{rigidify via growth}. 

Recent work has suggested that real monolayers of epithelial cells in curved space may adopt configurations in which the apical and basal surfaces of a cell have very different geometries \cite{rupprecht2017geometric,gomez2018scutoids,nelson2018epithelial}. Whether the apical vertex models considered in this and related works are still sufficiently expressive coarse grained models to capture the underlying physics requires further work; it may be that in curved space models written only in terms of a single cross-sectional plane are sufficient only when the individual cells are small enough to not appreciable feel the effects of curvature. A test of these models could be provided by ongoing experiments conducted on cellular monolayers on substrates with non-trivial curvature (e.g., Refs. \cite{harmand20203d,turiv2020topology}), as the present formulation could be readily extended from spherical constraints to more general ones (tori, Gaussian bumps, and simple sinusoidal profiles being particularly straightforward to implement numerically).

Before considering such complications, interesting extensions of the spherical vertex model presented here are anticipated by some existing studies of apical vertex models on curved surfaces \cite{Fletcher2014,osterfield2013three,murisic2015discrete}, in which the curved space is not a fixed embedding but can itself evolve and deform as the cells collectively exert stresses on their environment. It will be very interesting to combine models in which the surface can fluctuate rather than serving as a holonomic constraint with dynamical models that allow for cellular rearrangements.

Additionally, it will be very interesting to investigate the finite temperature glassy dynamics of this model in greater detail. Previous work on planar vertex and Voronoi models identified a deeply anomalous type of ``sub-Arrhenius'' dynamics, in which the relaxation time of the cells grew more slowly than exponential with decreasing temperature \cite{Sussman2018epl}. One speculation relates these unusual glassy dynamics to the unusual, residual-stress-driven rigidity transition those models possess at zero temperature. Embedding the vertex model on a sphere, as we have done here, provides one way to formally probe this hypothesis. We have constrained the vertices to lie exactly on the sphere, but have included no other energetic terms related to the curvature. Previous works on the apical vertex model have taken Eq. \ref{eq:energy1} and supplemented it with a discrete bending energy term \cite{seung1988defects} of the form
\begin{equation}\label{eq:bending}
E_b = B\sum_{ij}\left(1-\hat{n}_i\cdot \hat{n}_j\right),
\end{equation}
where $i$ and $j$ run over all neighboring faces and $\hat{n}_i$ is the surface normal corresponding to cell $i$.

Adding such a term is clearly relevant for the case where the surface can fluctuate, but it is \emph{also} interesting even for the perfectly spherical case. On a unit sphere note that the geodesic distance between two points is $|r_{ij}| = \cos^{-1}\left(\hat{n}_i\cdot \hat{n}_j\right)$; in the limit where the inter-cellular spacing is small compared to the radius of the sphere the above term can be approximated by $E_b \approx \frac{B}{2}\sum_{ij}r_{ij}^2,$ adding an additional quadratic constraint for ever pair of cellular neighbors. Whereas Eq. \ref{eq:energy1} represents an extensively underconstrained system that can only rigidify through residual stresses, Eq. \ref{eq:bending} introduces enough additional constraints to rigidify the system more conventionally. Thus, studying the glassy dynamics of the spherical vertex model as a function of tuning $B$ from zero to unity could test the root cause of the anomalous glassy behavior seen in other simple models of cellular matter. This would also further support our finding that the mechanism discussed here in the context of a particular underconstrained vertex model is present more generally in the class of space-filling or shape-based models of cellular matter.

In summary, in this work we have studied both the zero-temperature rigidity and the finite-temperature glassy dynamics of apical vertex models constrained to the surface of spheres, for which the energy functional is now expressed in terms of geodesic distances and the areas of spherical polygons. This constraint introduces a new ratio of length scales to the usual vertex model: the size of the sphere relative to the typical cell size. The critical shape index previously found to control the zero-temperature rigidity transition is affected by this ratio of length scales, and we find two modes by which cells could either \emph{rigidify} or \emph{unjam} as they divide, depending on whether the cellular division is accompanied by a growth of the spherical domain that keeps the cell cross-sectional area fixed or not. Finite-temperature studies show that the glassy dynamics of the spherical vertex model is sensitive to the underlying changes in the zero-temperature rigidity transition. Although the vertex model considered here is a somewhat specialized model of cellular monolayers, we emphasize that we expect the results obtained here -- which fundamentally stem from introducing curvature to the space in which the monolayer is constrained to move -- to be generic.

\begin{acknowledgments}
I would like to thank Matthias Merkel and Michael Moshe for illuminating discussions and for critical readings of this manuscript.
\end{acknowledgments}

\bibliography{sphericalVM}

\begin{thebibliography}{55}%
\makeatletter
\providecommand \@ifxundefined [1]{%
 \@ifx{#1\undefined}
}%
\providecommand \@ifnum [1]{%
 \ifnum #1\expandafter \@firstoftwo
 \else \expandafter \@secondoftwo
 \fi
}%
\providecommand \@ifx [1]{%
 \ifx #1\expandafter \@firstoftwo
 \else \expandafter \@secondoftwo
 \fi
}%
\providecommand \natexlab [1]{#1}%
\providecommand \enquote  [1]{``#1''}%
\providecommand \bibnamefont  [1]{#1}%
\providecommand \bibfnamefont [1]{#1}%
\providecommand \citenamefont [1]{#1}%
\providecommand \href@noop [0]{\@secondoftwo}%
\providecommand \href [0]{\begingroup \@sanitize@url \@href}%
\providecommand \@href[1]{\@@startlink{#1}\@@href}%
\providecommand \@@href[1]{\endgroup#1\@@endlink}%
\providecommand \@sanitize@url [0]{\catcode `\\12\catcode `\$12\catcode
  `\&12\catcode `\#12\catcode `\^12\catcode `\_12\catcode `\%12\relax}%
\providecommand \@@startlink[1]{}%
\providecommand \@@endlink[0]{}%
\providecommand \url  [0]{\begingroup\@sanitize@url \@url }%
\providecommand \@url [1]{\endgroup\@href {#1}{\urlprefix }}%
\providecommand \urlprefix  [0]{URL }%
\providecommand \Eprint [0]{\href }%
\providecommand \doibase [0]{http://dx.doi.org/}%
\providecommand \selectlanguage [0]{\@gobble}%
\providecommand \bibinfo  [0]{\@secondoftwo}%
\providecommand \bibfield  [0]{\@secondoftwo}%
\providecommand \translation [1]{[#1]}%
\providecommand \BibitemOpen [0]{}%
\providecommand \bibitemStop [0]{}%
\providecommand \bibitemNoStop [0]{.\EOS\space}%
\providecommand \EOS [0]{\spacefactor3000\relax}%
\providecommand \BibitemShut  [1]{\csname bibitem#1\endcsname}%
\let\auto@bib@innerbib\@empty
\bibitem [{\citenamefont {Thompson}(1952)}]{d1952growth}%
  \BibitemOpen
  \bibfield  {author} {\bibinfo {author} {\bibfnamefont {d.~W.}\ \bibnamefont
  {Thompson}},\ }\href@noop {} {\emph {\bibinfo {title} {On growth and
  form}}},\ Vol.~\bibinfo {volume} {1}\ (\bibinfo  {publisher} {Cambridge
  university press},\ \bibinfo {year} {1952})\BibitemShut {NoStop}%
\bibitem [{\citenamefont {Irvine}\ and\ \citenamefont
  {Shraiman}(2017)}]{irvine2017mechanical}%
  \BibitemOpen
  \bibfield  {author} {\bibinfo {author} {\bibfnamefont {K.~D.}\ \bibnamefont
  {Irvine}}\ and\ \bibinfo {author} {\bibfnamefont {B.~I.}\ \bibnamefont
  {Shraiman}},\ }\href@noop {} {\bibfield  {journal} {\bibinfo  {journal}
  {Development}\ }\textbf {\bibinfo {volume} {144}},\ \bibinfo {pages} {4238}
  (\bibinfo {year} {2017})}\BibitemShut {NoStop}%
\bibitem [{\citenamefont {Osterfield}\ \emph {et~al.}(2013)\citenamefont
  {Osterfield}, \citenamefont {Du}, \citenamefont {Sch{\"u}pbach},
  \citenamefont {Wieschaus},\ and\ \citenamefont
  {Shvartsman}}]{osterfield2013three}%
  \BibitemOpen
  \bibfield  {author} {\bibinfo {author} {\bibfnamefont {M.}~\bibnamefont
  {Osterfield}}, \bibinfo {author} {\bibfnamefont {X.}~\bibnamefont {Du}},
  \bibinfo {author} {\bibfnamefont {T.}~\bibnamefont {Sch{\"u}pbach}}, \bibinfo
  {author} {\bibfnamefont {E.}~\bibnamefont {Wieschaus}}, \ and\ \bibinfo
  {author} {\bibfnamefont {S.~Y.}\ \bibnamefont {Shvartsman}},\ }\href@noop {}
  {\bibfield  {journal} {\bibinfo  {journal} {Developmental cell}\ }\textbf
  {\bibinfo {volume} {24}},\ \bibinfo {pages} {400} (\bibinfo {year}
  {2013})}\BibitemShut {NoStop}%
\bibitem [{\citenamefont {Goodwin}\ \emph {et~al.}(2019)\citenamefont
  {Goodwin}, \citenamefont {Mao}, \citenamefont {Guyomar}, \citenamefont
  {Miller}, \citenamefont {Radisky}, \citenamefont {Ko{\v{s}}mrlj},\ and\
  \citenamefont {Nelson}}]{goodwin2019smooth}%
  \BibitemOpen
  \bibfield  {author} {\bibinfo {author} {\bibfnamefont {K.}~\bibnamefont
  {Goodwin}}, \bibinfo {author} {\bibfnamefont {S.}~\bibnamefont {Mao}},
  \bibinfo {author} {\bibfnamefont {T.}~\bibnamefont {Guyomar}}, \bibinfo
  {author} {\bibfnamefont {E.}~\bibnamefont {Miller}}, \bibinfo {author}
  {\bibfnamefont {D.~C.}\ \bibnamefont {Radisky}}, \bibinfo {author}
  {\bibfnamefont {A.}~\bibnamefont {Ko{\v{s}}mrlj}}, \ and\ \bibinfo {author}
  {\bibfnamefont {C.~M.}\ \bibnamefont {Nelson}},\ }\href@noop {} {\bibfield
  {journal} {\bibinfo  {journal} {Development}\ }\textbf {\bibinfo {volume}
  {146}} (\bibinfo {year} {2019})}\BibitemShut {NoStop}%
\bibitem [{\citenamefont {Rupprecht}\ \emph {et~al.}(2017)\citenamefont
  {Rupprecht}, \citenamefont {Ong}, \citenamefont {Yin}, \citenamefont {Huang},
  \citenamefont {Dinh}, \citenamefont {Singh}, \citenamefont {Zhang},
  \citenamefont {Yu},\ and\ \citenamefont {Saunders}}]{rupprecht2017geometric}%
  \BibitemOpen
  \bibfield  {author} {\bibinfo {author} {\bibfnamefont {J.-F.}\ \bibnamefont
  {Rupprecht}}, \bibinfo {author} {\bibfnamefont {K.~H.}\ \bibnamefont {Ong}},
  \bibinfo {author} {\bibfnamefont {J.}~\bibnamefont {Yin}}, \bibinfo {author}
  {\bibfnamefont {A.}~\bibnamefont {Huang}}, \bibinfo {author} {\bibfnamefont
  {H.-H.-Q.}\ \bibnamefont {Dinh}}, \bibinfo {author} {\bibfnamefont {A.~P.}\
  \bibnamefont {Singh}}, \bibinfo {author} {\bibfnamefont {S.}~\bibnamefont
  {Zhang}}, \bibinfo {author} {\bibfnamefont {W.}~\bibnamefont {Yu}}, \ and\
  \bibinfo {author} {\bibfnamefont {T.~E.}\ \bibnamefont {Saunders}},\
  }\href@noop {} {\bibfield  {journal} {\bibinfo  {journal} {Molecular biology
  of the cell}\ }\textbf {\bibinfo {volume} {28}},\ \bibinfo {pages} {3582}
  (\bibinfo {year} {2017})}\BibitemShut {NoStop}%
\bibitem [{\citenamefont {Streichan}\ \emph {et~al.}(2018)\citenamefont
  {Streichan}, \citenamefont {Lefebvre}, \citenamefont {Noll}, \citenamefont
  {Wieschaus},\ and\ \citenamefont {Shraiman}}]{streichan2018global}%
  \BibitemOpen
  \bibfield  {author} {\bibinfo {author} {\bibfnamefont {S.~J.}\ \bibnamefont
  {Streichan}}, \bibinfo {author} {\bibfnamefont {M.~F.}\ \bibnamefont
  {Lefebvre}}, \bibinfo {author} {\bibfnamefont {N.}~\bibnamefont {Noll}},
  \bibinfo {author} {\bibfnamefont {E.~F.}\ \bibnamefont {Wieschaus}}, \ and\
  \bibinfo {author} {\bibfnamefont {B.~I.}\ \bibnamefont {Shraiman}},\
  }\href@noop {} {\bibfield  {journal} {\bibinfo  {journal} {Elife}\ }\textbf
  {\bibinfo {volume} {7}},\ \bibinfo {pages} {e27454} (\bibinfo {year}
  {2018})}\BibitemShut {NoStop}%
\bibitem [{\citenamefont {Fouchard}\ \emph {et~al.}(2020)\citenamefont
  {Fouchard}, \citenamefont {Wyatt}, \citenamefont {Proag}, \citenamefont
  {Lisica}, \citenamefont {Khalilgharibi}, \citenamefont {Recho}, \citenamefont
  {Suzanne}, \citenamefont {Kabla},\ and\ \citenamefont
  {Charras}}]{fouchard2020curling}%
  \BibitemOpen
  \bibfield  {author} {\bibinfo {author} {\bibfnamefont {J.}~\bibnamefont
  {Fouchard}}, \bibinfo {author} {\bibfnamefont {T.~P.}\ \bibnamefont {Wyatt}},
  \bibinfo {author} {\bibfnamefont {A.}~\bibnamefont {Proag}}, \bibinfo
  {author} {\bibfnamefont {A.}~\bibnamefont {Lisica}}, \bibinfo {author}
  {\bibfnamefont {N.}~\bibnamefont {Khalilgharibi}}, \bibinfo {author}
  {\bibfnamefont {P.}~\bibnamefont {Recho}}, \bibinfo {author} {\bibfnamefont
  {M.}~\bibnamefont {Suzanne}}, \bibinfo {author} {\bibfnamefont
  {A.}~\bibnamefont {Kabla}}, \ and\ \bibinfo {author} {\bibfnamefont
  {G.}~\bibnamefont {Charras}},\ }\href@noop {} {\bibfield  {journal} {\bibinfo
   {journal} {Proceedings of the National Academy of Sciences}\ }\textbf
  {\bibinfo {volume} {117}},\ \bibinfo {pages} {9377} (\bibinfo {year}
  {2020})}\BibitemShut {NoStop}%
\bibitem [{\citenamefont {Brodland}(2004)}]{Brodland2004}%
  \BibitemOpen
  \bibfield  {author} {\bibinfo {author} {\bibfnamefont {G.~W.}\ \bibnamefont
  {Brodland}},\ }\href@noop {} {\bibfield  {journal} {\bibinfo  {journal}
  {Applied Mech. Rev.}\ }\textbf {\bibinfo {volume} {57}},\ \bibinfo {pages}
  {47} (\bibinfo {year} {2004})}\BibitemShut {NoStop}%
\bibitem [{\citenamefont {Van~Liedekerke}\ \emph {et~al.}(2015)\citenamefont
  {Van~Liedekerke}, \citenamefont {Palm}, \citenamefont {Jagiella},\ and\
  \citenamefont {Drasdo}}]{van2015simulating}%
  \BibitemOpen
  \bibfield  {author} {\bibinfo {author} {\bibfnamefont {P.}~\bibnamefont
  {Van~Liedekerke}}, \bibinfo {author} {\bibfnamefont {M.}~\bibnamefont
  {Palm}}, \bibinfo {author} {\bibfnamefont {N.}~\bibnamefont {Jagiella}}, \
  and\ \bibinfo {author} {\bibfnamefont {D.}~\bibnamefont {Drasdo}},\
  }\href@noop {} {\bibfield  {journal} {\bibinfo  {journal} {Computational
  particle mechanics}\ }\textbf {\bibinfo {volume} {2}},\ \bibinfo {pages}
  {401} (\bibinfo {year} {2015})}\BibitemShut {NoStop}%
\bibitem [{\citenamefont {Teomy}\ \emph {et~al.}(2018)\citenamefont {Teomy},
  \citenamefont {Kessler},\ and\ \citenamefont {Levine}}]{teomy2018confluent}%
  \BibitemOpen
  \bibfield  {author} {\bibinfo {author} {\bibfnamefont {E.}~\bibnamefont
  {Teomy}}, \bibinfo {author} {\bibfnamefont {D.~A.}\ \bibnamefont {Kessler}},
  \ and\ \bibinfo {author} {\bibfnamefont {H.}~\bibnamefont {Levine}},\
  }\href@noop {} {\bibfield  {journal} {\bibinfo  {journal} {Physical Review
  E}\ }\textbf {\bibinfo {volume} {98}},\ \bibinfo {pages} {042418} (\bibinfo
  {year} {2018})}\BibitemShut {NoStop}%
\bibitem [{\citenamefont {Boromand}\ \emph {et~al.}(2018)\citenamefont
  {Boromand}, \citenamefont {Signoriello}, \citenamefont {Ye}, \citenamefont
  {O'Hern},\ and\ \citenamefont {Shattuck}}]{boromand2018jamming}%
  \BibitemOpen
  \bibfield  {author} {\bibinfo {author} {\bibfnamefont {A.}~\bibnamefont
  {Boromand}}, \bibinfo {author} {\bibfnamefont {A.}~\bibnamefont
  {Signoriello}}, \bibinfo {author} {\bibfnamefont {F.}~\bibnamefont {Ye}},
  \bibinfo {author} {\bibfnamefont {C.~S.}\ \bibnamefont {O'Hern}}, \ and\
  \bibinfo {author} {\bibfnamefont {M.~D.}\ \bibnamefont {Shattuck}},\
  }\href@noop {} {\bibfield  {journal} {\bibinfo  {journal} {Physical review
  letters}\ }\textbf {\bibinfo {volume} {121}},\ \bibinfo {pages} {248003}
  (\bibinfo {year} {2018})}\BibitemShut {NoStop}%
\bibitem [{\citenamefont {Honda}(1978)}]{Honda1978}%
  \BibitemOpen
  \bibfield  {author} {\bibinfo {author} {\bibfnamefont {H.}~\bibnamefont
  {Honda}},\ }\href@noop {} {\bibfield  {journal} {\bibinfo  {journal} {J.
  Theor. Biol.}\ }\textbf {\bibinfo {volume} {72}},\ \bibinfo {pages} {523}
  (\bibinfo {year} {1978})}\BibitemShut {NoStop}%
\bibitem [{\citenamefont {Nagai}\ and\ \citenamefont
  {Honda}(2001)}]{Honda2001}%
  \BibitemOpen
  \bibfield  {author} {\bibinfo {author} {\bibfnamefont {T.}~\bibnamefont
  {Nagai}}\ and\ \bibinfo {author} {\bibfnamefont {H.}~\bibnamefont {Honda}},\
  }\href@noop {} {\bibfield  {journal} {\bibinfo  {journal} {Philosophical
  Magazine Part B}\ }\textbf {\bibinfo {volume} {81}},\ \bibinfo {pages} {699}
  (\bibinfo {year} {2001})}\BibitemShut {NoStop}%
\bibitem [{\citenamefont {Bi}\ \emph {et~al.}(2014)\citenamefont {Bi},
  \citenamefont {Lopez}, \citenamefont {Schwarz},\ and\ \citenamefont
  {Manning}}]{Bi2014}%
  \BibitemOpen
  \bibfield  {author} {\bibinfo {author} {\bibfnamefont {D.}~\bibnamefont
  {Bi}}, \bibinfo {author} {\bibfnamefont {J.~H.}\ \bibnamefont {Lopez}},
  \bibinfo {author} {\bibfnamefont {J.~M.}\ \bibnamefont {Schwarz}}, \ and\
  \bibinfo {author} {\bibfnamefont {M.~L.}\ \bibnamefont {Manning}},\
  }\href@noop {} {\bibfield  {journal} {\bibinfo  {journal} {Soft Matter}\
  }\textbf {\bibinfo {volume} {10}},\ \bibinfo {pages} {1885} (\bibinfo {year}
  {2014})}\BibitemShut {NoStop}%
\bibitem [{\citenamefont {Hufnagel}\ \emph {et~al.}(2007)\citenamefont
  {Hufnagel}, \citenamefont {Teleman}, \citenamefont {Rouault}, \citenamefont
  {Cohen},\ and\ \citenamefont {Shraiman}}]{Hufnagel2006}%
  \BibitemOpen
  \bibfield  {author} {\bibinfo {author} {\bibfnamefont {L.}~\bibnamefont
  {Hufnagel}}, \bibinfo {author} {\bibfnamefont {A.~A.}\ \bibnamefont
  {Teleman}}, \bibinfo {author} {\bibfnamefont {H.}~\bibnamefont {Rouault}},
  \bibinfo {author} {\bibfnamefont {S.~M.}\ \bibnamefont {Cohen}}, \ and\
  \bibinfo {author} {\bibfnamefont {B.~I.}\ \bibnamefont {Shraiman}},\
  }\href@noop {} {\bibfield  {journal} {\bibinfo  {journal} {Proc. Natl. Acad.
  Sci. USA}\ }\textbf {\bibinfo {volume} {104}},\ \bibinfo {pages} {3835}
  (\bibinfo {year} {2007})}\BibitemShut {NoStop}%
\bibitem [{\citenamefont {Farhadifar}\ \emph {et~al.}(2007)\citenamefont
  {Farhadifar}, \citenamefont {R{\"o}per}, \citenamefont {Aigouy},
  \citenamefont {Eaton},\ and\ \citenamefont {J{\"u}licher}}]{Farhadifar2007}%
  \BibitemOpen
  \bibfield  {author} {\bibinfo {author} {\bibfnamefont {R.}~\bibnamefont
  {Farhadifar}}, \bibinfo {author} {\bibfnamefont {J.-C.}\ \bibnamefont
  {R{\"o}per}}, \bibinfo {author} {\bibfnamefont {B.}~\bibnamefont {Aigouy}},
  \bibinfo {author} {\bibfnamefont {S.}~\bibnamefont {Eaton}}, \ and\ \bibinfo
  {author} {\bibfnamefont {F.}~\bibnamefont {J{\"u}licher}},\ }\href@noop {}
  {\bibfield  {journal} {\bibinfo  {journal} {Curr. Biol.}\ }\textbf {\bibinfo
  {volume} {17}},\ \bibinfo {pages} {2095} (\bibinfo {year}
  {2007})}\BibitemShut {NoStop}%
\bibitem [{\citenamefont {Friedl}\ and\ \citenamefont
  {Gilmour}(2009)}]{Friedl2009}%
  \BibitemOpen
  \bibfield  {author} {\bibinfo {author} {\bibfnamefont {P.}~\bibnamefont
  {Friedl}}\ and\ \bibinfo {author} {\bibfnamefont {D.}~\bibnamefont
  {Gilmour}},\ }\href@noop {} {\bibfield  {journal} {\bibinfo  {journal}
  {Nature reviews. Molecular cell biology}\ }\textbf {\bibinfo {volume} {10}},\
  \bibinfo {pages} {445} (\bibinfo {year} {2009})}\BibitemShut {NoStop}%
\bibitem [{\citenamefont {Brugu{\'{e}}s}\ \emph {et~al.}(2014)\citenamefont
  {Brugu{\'{e}}s}, \citenamefont {Anon}, \citenamefont {Conte}, \citenamefont
  {Veldhuis}, \citenamefont {Gupta}, \citenamefont {Colombelli}, \citenamefont
  {Mu{\~{n}}oz}, \citenamefont {Brodland}, \citenamefont {Ladoux},\ and\
  \citenamefont {Trepat}}]{Brugues2014}%
  \BibitemOpen
  \bibfield  {author} {\bibinfo {author} {\bibfnamefont {A.}~\bibnamefont
  {Brugu{\'{e}}s}}, \bibinfo {author} {\bibfnamefont {E.}~\bibnamefont {Anon}},
  \bibinfo {author} {\bibfnamefont {V.}~\bibnamefont {Conte}}, \bibinfo
  {author} {\bibfnamefont {J.~H.}\ \bibnamefont {Veldhuis}}, \bibinfo {author}
  {\bibfnamefont {M.}~\bibnamefont {Gupta}}, \bibinfo {author} {\bibfnamefont
  {J.}~\bibnamefont {Colombelli}}, \bibinfo {author} {\bibfnamefont {J.~J.}\
  \bibnamefont {Mu{\~{n}}oz}}, \bibinfo {author} {\bibfnamefont {G.~W.}\
  \bibnamefont {Brodland}}, \bibinfo {author} {\bibfnamefont {B.}~\bibnamefont
  {Ladoux}}, \ and\ \bibinfo {author} {\bibfnamefont {X.}~\bibnamefont
  {Trepat}},\ }\href@noop {} {\bibfield  {journal} {\bibinfo  {journal} {Nature
  Physics}\ }\textbf {\bibinfo {volume} {10}},\ \bibinfo {pages} {683}
  (\bibinfo {year} {2014})}\BibitemShut {NoStop}%
\bibitem [{\citenamefont {Etournay}\ \emph {et~al.}(2015)\citenamefont
  {Etournay}, \citenamefont {Popovi{\'{c}}}, \citenamefont {Merkel},
  \citenamefont {Nandi}, \citenamefont {Blasse}, \citenamefont {Aigouy},
  \citenamefont {Brandl}, \citenamefont {Myers}, \citenamefont {Salbreux},
  \citenamefont {J{\"{u}}licher},\ and\ \citenamefont {Eaton}}]{Etournay2015}%
  \BibitemOpen
  \bibfield  {author} {\bibinfo {author} {\bibfnamefont {R.}~\bibnamefont
  {Etournay}}, \bibinfo {author} {\bibfnamefont {M.}~\bibnamefont
  {Popovi{\'{c}}}}, \bibinfo {author} {\bibfnamefont {M.}~\bibnamefont
  {Merkel}}, \bibinfo {author} {\bibfnamefont {A.}~\bibnamefont {Nandi}},
  \bibinfo {author} {\bibfnamefont {C.}~\bibnamefont {Blasse}}, \bibinfo
  {author} {\bibfnamefont {B.}~\bibnamefont {Aigouy}}, \bibinfo {author}
  {\bibfnamefont {H.}~\bibnamefont {Brandl}}, \bibinfo {author} {\bibfnamefont
  {G.}~\bibnamefont {Myers}}, \bibinfo {author} {\bibfnamefont
  {G.}~\bibnamefont {Salbreux}}, \bibinfo {author} {\bibfnamefont
  {F.}~\bibnamefont {J{\"{u}}licher}}, \ and\ \bibinfo {author} {\bibfnamefont
  {S.}~\bibnamefont {Eaton}},\ }\href@noop {} {\bibfield  {journal} {\bibinfo
  {journal} {eLife}\ }\textbf {\bibinfo {volume} {4}},\ \bibinfo {pages}
  {e07090} (\bibinfo {year} {2015})}\BibitemShut {NoStop}%
\bibitem [{\citenamefont {Spahn}\ and\ \citenamefont
  {Reuter}(2013)}]{spahn2013vertex}%
  \BibitemOpen
  \bibfield  {author} {\bibinfo {author} {\bibfnamefont {P.}~\bibnamefont
  {Spahn}}\ and\ \bibinfo {author} {\bibfnamefont {R.}~\bibnamefont {Reuter}},\
  }\href@noop {} {\bibfield  {journal} {\bibinfo  {journal} {PloS one}\
  }\textbf {\bibinfo {volume} {8}} (\bibinfo {year} {2013})}\BibitemShut
  {NoStop}%
\bibitem [{\citenamefont {Staple}\ \emph {et~al.}(2010)\citenamefont {Staple},
  \citenamefont {Farhadifar}, \citenamefont {R{\"o}per}, \citenamefont
  {Aigouy}, \citenamefont {Eaton},\ and\ \citenamefont
  {J{\"u}licher}}]{staple2010mechanics}%
  \BibitemOpen
  \bibfield  {author} {\bibinfo {author} {\bibfnamefont {D.~B.}\ \bibnamefont
  {Staple}}, \bibinfo {author} {\bibfnamefont {R.}~\bibnamefont {Farhadifar}},
  \bibinfo {author} {\bibfnamefont {J.-C.}\ \bibnamefont {R{\"o}per}}, \bibinfo
  {author} {\bibfnamefont {B.}~\bibnamefont {Aigouy}}, \bibinfo {author}
  {\bibfnamefont {S.}~\bibnamefont {Eaton}}, \ and\ \bibinfo {author}
  {\bibfnamefont {F.}~\bibnamefont {J{\"u}licher}},\ }\href@noop {} {\bibfield
  {journal} {\bibinfo  {journal} {The European Physical Journal E}\ }\textbf
  {\bibinfo {volume} {33}},\ \bibinfo {pages} {117} (\bibinfo {year}
  {2010})}\BibitemShut {NoStop}%
\bibitem [{\citenamefont {Bi}\ \emph {et~al.}(2015)\citenamefont {Bi},
  \citenamefont {Lopez}, \citenamefont {Schwarz},\ and\ \citenamefont
  {Manning}}]{Bi2015}%
  \BibitemOpen
  \bibfield  {author} {\bibinfo {author} {\bibfnamefont {D.}~\bibnamefont
  {Bi}}, \bibinfo {author} {\bibfnamefont {J.}~\bibnamefont {Lopez}}, \bibinfo
  {author} {\bibfnamefont {J.}~\bibnamefont {Schwarz}}, \ and\ \bibinfo
  {author} {\bibfnamefont {M.~L.}\ \bibnamefont {Manning}},\ }\href@noop {}
  {\bibfield  {journal} {\bibinfo  {journal} {Nature Physics}\ }\textbf
  {\bibinfo {volume} {11}},\ \bibinfo {pages} {1074} (\bibinfo {year}
  {2015})}\BibitemShut {NoStop}%
\bibitem [{\citenamefont {Sussman}\ and\ \citenamefont
  {Merkel}(2018)}]{sussman2018no}%
  \BibitemOpen
  \bibfield  {author} {\bibinfo {author} {\bibfnamefont {D.~M.}\ \bibnamefont
  {Sussman}}\ and\ \bibinfo {author} {\bibfnamefont {M.}~\bibnamefont
  {Merkel}},\ }\href@noop {} {\bibfield  {journal} {\bibinfo  {journal} {Soft
  matter}\ }\textbf {\bibinfo {volume} {14}},\ \bibinfo {pages} {3397}
  (\bibinfo {year} {2018})}\BibitemShut {NoStop}%
\bibitem [{\citenamefont {Yan}\ and\ \citenamefont
  {Bi}(2019)}]{yan2019multicellular}%
  \BibitemOpen
  \bibfield  {author} {\bibinfo {author} {\bibfnamefont {L.}~\bibnamefont
  {Yan}}\ and\ \bibinfo {author} {\bibfnamefont {D.}~\bibnamefont {Bi}},\
  }\href@noop {} {\bibfield  {journal} {\bibinfo  {journal} {Physical Review
  X}\ }\textbf {\bibinfo {volume} {9}},\ \bibinfo {pages} {011029} (\bibinfo
  {year} {2019})}\BibitemShut {NoStop}%
\bibitem [{\citenamefont {Moshe}\ \emph {et~al.}(2018)\citenamefont {Moshe},
  \citenamefont {Bowick},\ and\ \citenamefont
  {Marchetti}}]{moshe2018geometric}%
  \BibitemOpen
  \bibfield  {author} {\bibinfo {author} {\bibfnamefont {M.}~\bibnamefont
  {Moshe}}, \bibinfo {author} {\bibfnamefont {M.~J.}\ \bibnamefont {Bowick}}, \
  and\ \bibinfo {author} {\bibfnamefont {M.~C.}\ \bibnamefont {Marchetti}},\
  }\href@noop {} {\bibfield  {journal} {\bibinfo  {journal} {Physical review
  letters}\ }\textbf {\bibinfo {volume} {120}},\ \bibinfo {pages} {268105}
  (\bibinfo {year} {2018})}\BibitemShut {NoStop}%
\bibitem [{\citenamefont {Noll}\ \emph {et~al.}(2017)\citenamefont {Noll},
  \citenamefont {Mani}, \citenamefont {Heemskerk}, \citenamefont {Streichan},\
  and\ \citenamefont {Shraiman}}]{noll2017active}%
  \BibitemOpen
  \bibfield  {author} {\bibinfo {author} {\bibfnamefont {N.}~\bibnamefont
  {Noll}}, \bibinfo {author} {\bibfnamefont {M.}~\bibnamefont {Mani}}, \bibinfo
  {author} {\bibfnamefont {I.}~\bibnamefont {Heemskerk}}, \bibinfo {author}
  {\bibfnamefont {S.~J.}\ \bibnamefont {Streichan}}, \ and\ \bibinfo {author}
  {\bibfnamefont {B.~I.}\ \bibnamefont {Shraiman}},\ }\href@noop {} {\bibfield
  {journal} {\bibinfo  {journal} {Nature Physics}\ }\textbf {\bibinfo {volume}
  {13}},\ \bibinfo {pages} {1221} (\bibinfo {year} {2017})}\BibitemShut
  {NoStop}%
\bibitem [{\citenamefont {Sussman}\ \emph
  {et~al.}(2018{\natexlab{a}})\citenamefont {Sussman}, \citenamefont
  {Paoluzzi}, \citenamefont {{Cristina Marchetti}},\ and\ \citenamefont {{Lisa
  Manning}}}]{Sussman2018epl}%
  \BibitemOpen
  \bibfield  {author} {\bibinfo {author} {\bibfnamefont {D.~M.}\ \bibnamefont
  {Sussman}}, \bibinfo {author} {\bibfnamefont {M.}~\bibnamefont {Paoluzzi}},
  \bibinfo {author} {\bibfnamefont {M.}~\bibnamefont {{Cristina Marchetti}}}, \
  and\ \bibinfo {author} {\bibfnamefont {M.}~\bibnamefont {{Lisa Manning}}},\
  }\href@noop {} {\bibfield  {journal} {\bibinfo  {journal} {EPL (Europhysics
  Letters)}\ }\textbf {\bibinfo {volume} {121}},\ \bibinfo {pages} {36001}
  (\bibinfo {year} {2018}{\natexlab{a}})}\BibitemShut {NoStop}%
\bibitem [{\citenamefont {Sussman}\ \emph
  {et~al.}(2018{\natexlab{b}})\citenamefont {Sussman}, \citenamefont {Schwarz},
  \citenamefont {Marchetti},\ and\ \citenamefont {Manning}}]{sussman2018soft}%
  \BibitemOpen
  \bibfield  {author} {\bibinfo {author} {\bibfnamefont {D.~M.}\ \bibnamefont
  {Sussman}}, \bibinfo {author} {\bibfnamefont {J.~M.}\ \bibnamefont
  {Schwarz}}, \bibinfo {author} {\bibfnamefont {M.~C.}\ \bibnamefont
  {Marchetti}}, \ and\ \bibinfo {author} {\bibfnamefont {M.~L.}\ \bibnamefont
  {Manning}},\ }\href@noop {} {\bibfield  {journal} {\bibinfo  {journal}
  {Physical review letters}\ }\textbf {\bibinfo {volume} {120}},\ \bibinfo
  {pages} {058001} (\bibinfo {year} {2018}{\natexlab{b}})}\BibitemShut
  {NoStop}%
\bibitem [{\citenamefont {Alt}\ \emph {et~al.}(2017)\citenamefont {Alt},
  \citenamefont {Ganguly},\ and\ \citenamefont {Salbreux}}]{alt2017vertex}%
  \BibitemOpen
  \bibfield  {author} {\bibinfo {author} {\bibfnamefont {S.}~\bibnamefont
  {Alt}}, \bibinfo {author} {\bibfnamefont {P.}~\bibnamefont {Ganguly}}, \ and\
  \bibinfo {author} {\bibfnamefont {G.}~\bibnamefont {Salbreux}},\ }\href@noop
  {} {\bibfield  {journal} {\bibinfo  {journal} {Philosophical Transactions of
  the Royal Society B: Biological Sciences}\ }\textbf {\bibinfo {volume}
  {372}},\ \bibinfo {pages} {20150520} (\bibinfo {year} {2017})}\BibitemShut
  {NoStop}%
\bibitem [{\citenamefont {Okuda}\ \emph {et~al.}(2015)\citenamefont {Okuda},
  \citenamefont {Inoue},\ and\ \citenamefont {Adachi}}]{okuda2015three}%
  \BibitemOpen
  \bibfield  {author} {\bibinfo {author} {\bibfnamefont {S.}~\bibnamefont
  {Okuda}}, \bibinfo {author} {\bibfnamefont {Y.}~\bibnamefont {Inoue}}, \ and\
  \bibinfo {author} {\bibfnamefont {T.}~\bibnamefont {Adachi}},\ }\href@noop {}
  {\bibfield  {journal} {\bibinfo  {journal} {Biophysics and physicobiology}\
  }\textbf {\bibinfo {volume} {12}},\ \bibinfo {pages} {13} (\bibinfo {year}
  {2015})}\BibitemShut {NoStop}%
\bibitem [{\citenamefont {Fletcher}\ \emph {et~al.}(2014)\citenamefont
  {Fletcher}, \citenamefont {Osterfield}, \citenamefont {Baker},\ and\
  \citenamefont {Shvartsman}}]{Fletcher2014}%
  \BibitemOpen
  \bibfield  {author} {\bibinfo {author} {\bibfnamefont {A.~G.}\ \bibnamefont
  {Fletcher}}, \bibinfo {author} {\bibfnamefont {M.}~\bibnamefont
  {Osterfield}}, \bibinfo {author} {\bibfnamefont {R.~E.}\ \bibnamefont
  {Baker}}, \ and\ \bibinfo {author} {\bibfnamefont {S.~Y.}\ \bibnamefont
  {Shvartsman}},\ }\href@noop {} {\bibfield  {journal} {\bibinfo  {journal}
  {Biophys. J.}\ }\textbf {\bibinfo {volume} {106}},\ \bibinfo {pages} {2291}
  (\bibinfo {year} {2014})}\BibitemShut {NoStop}%
\bibitem [{\citenamefont {Murisic}\ \emph {et~al.}(2015)\citenamefont
  {Murisic}, \citenamefont {Hakim}, \citenamefont {Kevrekidis}, \citenamefont
  {Shvartsman},\ and\ \citenamefont {Audoly}}]{murisic2015discrete}%
  \BibitemOpen
  \bibfield  {author} {\bibinfo {author} {\bibfnamefont {N.}~\bibnamefont
  {Murisic}}, \bibinfo {author} {\bibfnamefont {V.}~\bibnamefont {Hakim}},
  \bibinfo {author} {\bibfnamefont {I.~G.}\ \bibnamefont {Kevrekidis}},
  \bibinfo {author} {\bibfnamefont {S.~Y.}\ \bibnamefont {Shvartsman}}, \ and\
  \bibinfo {author} {\bibfnamefont {B.}~\bibnamefont {Audoly}},\ }\href@noop {}
  {\bibfield  {journal} {\bibinfo  {journal} {Biophysical journal}\ }\textbf
  {\bibinfo {volume} {109}},\ \bibinfo {pages} {154} (\bibinfo {year}
  {2015})}\BibitemShut {NoStop}%
\bibitem [{\citenamefont {Barton}\ \emph {et~al.}(2017)\citenamefont {Barton},
  \citenamefont {Henkes}, \citenamefont {Weijer},\ and\ \citenamefont
  {Sknepnek}}]{SAMOS}%
  \BibitemOpen
  \bibfield  {author} {\bibinfo {author} {\bibfnamefont {D.~L.}\ \bibnamefont
  {Barton}}, \bibinfo {author} {\bibfnamefont {S.}~\bibnamefont {Henkes}},
  \bibinfo {author} {\bibfnamefont {C.~J.}\ \bibnamefont {Weijer}}, \ and\
  \bibinfo {author} {\bibfnamefont {R.}~\bibnamefont {Sknepnek}},\ }\href@noop
  {} {\bibfield  {journal} {\bibinfo  {journal} {PLoS Computational Biology}\
  }\textbf {\bibinfo {volume} {13}},\ \bibinfo {pages} {e1005569} (\bibinfo
  {year} {2017})}\BibitemShut {NoStop}%
\bibitem [{\citenamefont {Kim}\ \emph {et~al.}(2018)\citenamefont {Kim},
  \citenamefont {Wang},\ and\ \citenamefont {Hilgenfeldt}}]{kim2018universal}%
  \BibitemOpen
  \bibfield  {author} {\bibinfo {author} {\bibfnamefont {S.}~\bibnamefont
  {Kim}}, \bibinfo {author} {\bibfnamefont {Y.}~\bibnamefont {Wang}}, \ and\
  \bibinfo {author} {\bibfnamefont {S.}~\bibnamefont {Hilgenfeldt}},\
  }\href@noop {} {\bibfield  {journal} {\bibinfo  {journal} {Physical review
  letters}\ }\textbf {\bibinfo {volume} {120}},\ \bibinfo {pages} {248001}
  (\bibinfo {year} {2018})}\BibitemShut {NoStop}%
\bibitem [{\citenamefont {Merkel}\ and\ \citenamefont
  {Manning}(2018)}]{merkel2018geometrically}%
  \BibitemOpen
  \bibfield  {author} {\bibinfo {author} {\bibfnamefont {M.}~\bibnamefont
  {Merkel}}\ and\ \bibinfo {author} {\bibfnamefont {M.~L.}\ \bibnamefont
  {Manning}},\ }\href@noop {} {\bibfield  {journal} {\bibinfo  {journal} {New
  Journal of Physics}\ }\textbf {\bibinfo {volume} {20}},\ \bibinfo {pages}
  {022002} (\bibinfo {year} {2018})}\BibitemShut {NoStop}%
\bibitem [{\citenamefont {Tarjus}\ \emph {et~al.}(2012)\citenamefont {Tarjus},
  \citenamefont {Sausset},\ and\ \citenamefont {Viot}}]{tarjus2012statistical}%
  \BibitemOpen
  \bibfield  {author} {\bibinfo {author} {\bibfnamefont {G.}~\bibnamefont
  {Tarjus}}, \bibinfo {author} {\bibfnamefont {E.}~\bibnamefont {Sausset}}, \
  and\ \bibinfo {author} {\bibfnamefont {P.}~\bibnamefont {Viot}},\ }\href@noop
  {} {\bibfield  {journal} {\bibinfo  {journal} {Advances in Chemical Physics}\
  }\textbf {\bibinfo {volume} {148}},\ \bibinfo {pages} {251} (\bibinfo {year}
  {2012})}\BibitemShut {NoStop}%
\bibitem [{\citenamefont {Guerra}\ \emph {et~al.}(2018)\citenamefont {Guerra},
  \citenamefont {Kelleher}, \citenamefont {Hollingsworth},\ and\ \citenamefont
  {Chaikin}}]{guerra2018freezing}%
  \BibitemOpen
  \bibfield  {author} {\bibinfo {author} {\bibfnamefont {R.~E.}\ \bibnamefont
  {Guerra}}, \bibinfo {author} {\bibfnamefont {C.~P.}\ \bibnamefont
  {Kelleher}}, \bibinfo {author} {\bibfnamefont {A.~D.}\ \bibnamefont
  {Hollingsworth}}, \ and\ \bibinfo {author} {\bibfnamefont {P.~M.}\
  \bibnamefont {Chaikin}},\ }\href@noop {} {\bibfield  {journal} {\bibinfo
  {journal} {Nature}\ }\textbf {\bibinfo {volume} {554}},\ \bibinfo {pages}
  {346} (\bibinfo {year} {2018})}\BibitemShut {NoStop}%
\bibitem [{\citenamefont {Sussman}(2017)}]{sussman2017cellGPU}%
  \BibitemOpen
  \bibfield  {author} {\bibinfo {author} {\bibfnamefont {D.~M.}\ \bibnamefont
  {Sussman}},\ }\href@noop {} {\bibfield  {journal} {\bibinfo  {journal}
  {Comput. Phys. Commun.}\ }\textbf {\bibinfo {volume} {219}},\ \bibinfo
  {pages} {400} (\bibinfo {year} {2017})}\BibitemShut {NoStop}%
\bibitem [{\citenamefont {Sussman}\ and\ \citenamefont
  {Beller}(2019)}]{sussman2019fast}%
  \BibitemOpen
  \bibfield  {author} {\bibinfo {author} {\bibfnamefont {D.~M.}\ \bibnamefont
  {Sussman}}\ and\ \bibinfo {author} {\bibfnamefont {D.~A.}\ \bibnamefont
  {Beller}},\ }\href@noop {} {\bibfield  {journal} {\bibinfo  {journal}
  {Frontiers in Physics}\ }\textbf {\bibinfo {volume} {7}},\ \bibinfo {pages}
  {204} (\bibinfo {year} {2019})}\BibitemShut {NoStop}%
\bibitem [{\citenamefont {Sknepnek}\ and\ \citenamefont
  {Henkes}(2015)}]{sknepnek2015active}%
  \BibitemOpen
  \bibfield  {author} {\bibinfo {author} {\bibfnamefont {R.}~\bibnamefont
  {Sknepnek}}\ and\ \bibinfo {author} {\bibfnamefont {S.}~\bibnamefont
  {Henkes}},\ }\href@noop {} {\bibfield  {journal} {\bibinfo  {journal}
  {Physical Review E}\ }\textbf {\bibinfo {volume} {91}},\ \bibinfo {pages}
  {022306} (\bibinfo {year} {2015})}\BibitemShut {NoStop}%
\bibitem [{\citenamefont {Bi}\ \emph {et~al.}(2016)\citenamefont {Bi},
  \citenamefont {Yang}, \citenamefont {Marchetti},\ and\ \citenamefont
  {Manning}}]{bi2016motility}%
  \BibitemOpen
  \bibfield  {author} {\bibinfo {author} {\bibfnamefont {D.}~\bibnamefont
  {Bi}}, \bibinfo {author} {\bibfnamefont {X.}~\bibnamefont {Yang}}, \bibinfo
  {author} {\bibfnamefont {M.~C.}\ \bibnamefont {Marchetti}}, \ and\ \bibinfo
  {author} {\bibfnamefont {M.~L.}\ \bibnamefont {Manning}},\ }\href@noop {}
  {\bibfield  {journal} {\bibinfo  {journal} {Physical Review X}\ }\textbf
  {\bibinfo {volume} {6}},\ \bibinfo {pages} {021011} (\bibinfo {year}
  {2016})}\BibitemShut {NoStop}%
\bibitem [{\citenamefont {{The CGAL Project}}(2019)}]{cgal:eb-19b}%
  \BibitemOpen
  \bibfield  {author} {\bibinfo {author} {\bibnamefont {{The CGAL Project}}},\
  }\href {https://doc.cgal.org/5.0/Manual/packages.html} {\emph {\bibinfo
  {title} {{CGAL} User and Reference Manual}}},\ \bibinfo {edition} {{5.0}}\
  ed.\ (\bibinfo  {publisher} {{CGAL Editorial Board}},\ \bibinfo {year}
  {2019})\BibitemShut {NoStop}%
\bibitem [{\citenamefont {Hert}\ and\ \citenamefont
  {Schirra}(2019)}]{cgal:hs-ch3-19b}%
  \BibitemOpen
  \bibfield  {author} {\bibinfo {author} {\bibfnamefont {S.}~\bibnamefont
  {Hert}}\ and\ \bibinfo {author} {\bibfnamefont {S.}~\bibnamefont {Schirra}},\
  }in\ \href {https://doc.cgal.org/5.0/Manual/packages.html#PkgConvexHull3}
  {\emph {\bibinfo {booktitle} {{CGAL} User and Reference Manual}}}\ (\bibinfo
  {publisher} {{CGAL Editorial Board}},\ \bibinfo {year} {2019})\ \bibinfo
  {edition} {{5.0}}\ ed.\BibitemShut {Stop}%
\bibitem [{\citenamefont {Fortune}(1995)}]{fortune1995voronoi}%
  \BibitemOpen
  \bibfield  {author} {\bibinfo {author} {\bibfnamefont {S.}~\bibnamefont
  {Fortune}},\ }in\ \href@noop {} {\emph {\bibinfo {booktitle} {Computing in
  Euclidean geometry}}}\ (\bibinfo  {publisher} {World Scientific},\ \bibinfo
  {year} {1995})\ pp.\ \bibinfo {pages} {225--265}\BibitemShut {NoStop}%
\bibitem [{\citenamefont {Bitzek}\ \emph {et~al.}(2006)\citenamefont {Bitzek},
  \citenamefont {Koskinen}, \citenamefont {G{\"a}hler}, \citenamefont
  {Moseler},\ and\ \citenamefont {Gumbsch}}]{bitzek2006structural}%
  \BibitemOpen
  \bibfield  {author} {\bibinfo {author} {\bibfnamefont {E.}~\bibnamefont
  {Bitzek}}, \bibinfo {author} {\bibfnamefont {P.}~\bibnamefont {Koskinen}},
  \bibinfo {author} {\bibfnamefont {F.}~\bibnamefont {G{\"a}hler}}, \bibinfo
  {author} {\bibfnamefont {M.}~\bibnamefont {Moseler}}, \ and\ \bibinfo
  {author} {\bibfnamefont {P.}~\bibnamefont {Gumbsch}},\ }\href@noop {}
  {\bibfield  {journal} {\bibinfo  {journal} {Physical review letters}\
  }\textbf {\bibinfo {volume} {97}},\ \bibinfo {pages} {170201} (\bibinfo
  {year} {2006})}\BibitemShut {NoStop}%
\bibitem [{\citenamefont {Merkel}\ \emph {et~al.}(2019)\citenamefont {Merkel},
  \citenamefont {Baumgarten}, \citenamefont {Tighe},\ and\ \citenamefont
  {Manning}}]{merkel2019minimal}%
  \BibitemOpen
  \bibfield  {author} {\bibinfo {author} {\bibfnamefont {M.}~\bibnamefont
  {Merkel}}, \bibinfo {author} {\bibfnamefont {K.}~\bibnamefont {Baumgarten}},
  \bibinfo {author} {\bibfnamefont {B.~P.}\ \bibnamefont {Tighe}}, \ and\
  \bibinfo {author} {\bibfnamefont {M.~L.}\ \bibnamefont {Manning}},\
  }\href@noop {} {\bibfield  {journal} {\bibinfo  {journal} {Proceedings of the
  National Academy of Sciences}\ }\textbf {\bibinfo {volume} {116}},\ \bibinfo
  {pages} {6560} (\bibinfo {year} {2019})}\BibitemShut {NoStop}%
\bibitem [{\citenamefont {Brakke}(1992)}]{brakke1992surface}%
  \BibitemOpen
  \bibfield  {author} {\bibinfo {author} {\bibfnamefont {K.~A.}\ \bibnamefont
  {Brakke}},\ }\href@noop {} {\bibfield  {journal} {\bibinfo  {journal}
  {Experimental mathematics}\ }\textbf {\bibinfo {volume} {1}},\ \bibinfo
  {pages} {141} (\bibinfo {year} {1992})}\BibitemShut {NoStop}%
\bibitem [{\citenamefont {Keys}\ \emph {et~al.}(2007)\citenamefont {Keys},
  \citenamefont {Abate}, \citenamefont {Glotzer},\ and\ \citenamefont
  {Durian}}]{keys2007measurement}%
  \BibitemOpen
  \bibfield  {author} {\bibinfo {author} {\bibfnamefont {A.~S.}\ \bibnamefont
  {Keys}}, \bibinfo {author} {\bibfnamefont {A.~R.}\ \bibnamefont {Abate}},
  \bibinfo {author} {\bibfnamefont {S.~C.}\ \bibnamefont {Glotzer}}, \ and\
  \bibinfo {author} {\bibfnamefont {D.~J.}\ \bibnamefont {Durian}},\
  }\href@noop {} {\bibfield  {journal} {\bibinfo  {journal} {Nature physics}\
  }\textbf {\bibinfo {volume} {3}},\ \bibinfo {pages} {260} (\bibinfo {year}
  {2007})}\BibitemShut {NoStop}%
\bibitem [{\citenamefont {Mongera}\ \emph {et~al.}(2018)\citenamefont
  {Mongera}, \citenamefont {Rowghanian}, \citenamefont {Gustafson},
  \citenamefont {Shelton}, \citenamefont {Kealhofer}, \citenamefont {Carn},
  \citenamefont {Serwane}, \citenamefont {Lucio}, \citenamefont {Giammona},\
  and\ \citenamefont {Camp{\`a}s}}]{mongera2018fluid}%
  \BibitemOpen
  \bibfield  {author} {\bibinfo {author} {\bibfnamefont {A.}~\bibnamefont
  {Mongera}}, \bibinfo {author} {\bibfnamefont {P.}~\bibnamefont {Rowghanian}},
  \bibinfo {author} {\bibfnamefont {H.~J.}\ \bibnamefont {Gustafson}}, \bibinfo
  {author} {\bibfnamefont {E.}~\bibnamefont {Shelton}}, \bibinfo {author}
  {\bibfnamefont {D.~A.}\ \bibnamefont {Kealhofer}}, \bibinfo {author}
  {\bibfnamefont {E.~K.}\ \bibnamefont {Carn}}, \bibinfo {author}
  {\bibfnamefont {F.}~\bibnamefont {Serwane}}, \bibinfo {author} {\bibfnamefont
  {A.~A.}\ \bibnamefont {Lucio}}, \bibinfo {author} {\bibfnamefont
  {J.}~\bibnamefont {Giammona}}, \ and\ \bibinfo {author} {\bibfnamefont
  {O.}~\bibnamefont {Camp{\`a}s}},\ }\href@noop {} {\bibfield  {journal}
  {\bibinfo  {journal} {Nature}\ }\textbf {\bibinfo {volume} {561}},\ \bibinfo
  {pages} {401} (\bibinfo {year} {2018})}\BibitemShut {NoStop}%
\bibitem [{\citenamefont {Jain}\ \emph {et~al.}(2019)\citenamefont {Jain},
  \citenamefont {Ulman}, \citenamefont {Mukherjee}, \citenamefont {Prakash},
  \citenamefont {Pimpale}, \citenamefont {M{\"u}nster}, \citenamefont
  {Panfilio}, \citenamefont {Jug}, \citenamefont {Grill}, \citenamefont
  {Tomancak} \emph {et~al.}}]{jain2019regionalized}%
  \BibitemOpen
  \bibfield  {author} {\bibinfo {author} {\bibfnamefont {A.}~\bibnamefont
  {Jain}}, \bibinfo {author} {\bibfnamefont {V.}~\bibnamefont {Ulman}},
  \bibinfo {author} {\bibfnamefont {A.}~\bibnamefont {Mukherjee}}, \bibinfo
  {author} {\bibfnamefont {M.}~\bibnamefont {Prakash}}, \bibinfo {author}
  {\bibfnamefont {L.}~\bibnamefont {Pimpale}}, \bibinfo {author} {\bibfnamefont
  {S.}~\bibnamefont {M{\"u}nster}}, \bibinfo {author} {\bibfnamefont {K.~A.}\
  \bibnamefont {Panfilio}}, \bibinfo {author} {\bibfnamefont {F.}~\bibnamefont
  {Jug}}, \bibinfo {author} {\bibfnamefont {S.~W.}\ \bibnamefont {Grill}},
  \bibinfo {author} {\bibfnamefont {P.}~\bibnamefont {Tomancak}},  \emph
  {et~al.},\ }\href@noop {} {\bibfield  {journal} {\bibinfo  {journal}
  {BioRxiv}\ ,\ \bibinfo {pages} {744193}} (\bibinfo {year}
  {2019})}\BibitemShut {NoStop}%
\bibitem [{\citenamefont {G{\'o}mez-G{\'a}lvez}\ \emph
  {et~al.}(2018)\citenamefont {G{\'o}mez-G{\'a}lvez}, \citenamefont
  {Vicente-Munuera}, \citenamefont {Tagua}, \citenamefont {Forja},
  \citenamefont {Castro}, \citenamefont {Letr{\'a}n}, \citenamefont
  {Valencia-Exp{\'o}sito}, \citenamefont {Grima}, \citenamefont
  {Berm{\'u}dez-Gallardo}, \citenamefont {Serrano-P{\'e}rez-Higueras} \emph
  {et~al.}}]{gomez2018scutoids}%
  \BibitemOpen
  \bibfield  {author} {\bibinfo {author} {\bibfnamefont {P.}~\bibnamefont
  {G{\'o}mez-G{\'a}lvez}}, \bibinfo {author} {\bibfnamefont {P.}~\bibnamefont
  {Vicente-Munuera}}, \bibinfo {author} {\bibfnamefont {A.}~\bibnamefont
  {Tagua}}, \bibinfo {author} {\bibfnamefont {C.}~\bibnamefont {Forja}},
  \bibinfo {author} {\bibfnamefont {A.~M.}\ \bibnamefont {Castro}}, \bibinfo
  {author} {\bibfnamefont {M.}~\bibnamefont {Letr{\'a}n}}, \bibinfo {author}
  {\bibfnamefont {A.}~\bibnamefont {Valencia-Exp{\'o}sito}}, \bibinfo {author}
  {\bibfnamefont {C.}~\bibnamefont {Grima}}, \bibinfo {author} {\bibfnamefont
  {M.}~\bibnamefont {Berm{\'u}dez-Gallardo}}, \bibinfo {author} {\bibfnamefont
  {{\'O}.}~\bibnamefont {Serrano-P{\'e}rez-Higueras}},  \emph {et~al.},\
  }\href@noop {} {\bibfield  {journal} {\bibinfo  {journal} {Nature
  communications}\ }\textbf {\bibinfo {volume} {9}},\ \bibinfo {pages} {2960}
  (\bibinfo {year} {2018})}\BibitemShut {NoStop}%
\bibitem [{\citenamefont {Nelson}(2018)}]{nelson2018epithelial}%
  \BibitemOpen
  \bibfield  {author} {\bibinfo {author} {\bibfnamefont {C.~M.}\ \bibnamefont
  {Nelson}},\ }\href@noop {} {\bibfield  {journal} {\bibinfo  {journal}
  {Current Biology}\ }\textbf {\bibinfo {volume} {28}},\ \bibinfo {pages}
  {R1197} (\bibinfo {year} {2018})}\BibitemShut {NoStop}%
\bibitem [{\citenamefont {Harmand}\ and\ \citenamefont
  {H{\'e}non}(2020)}]{harmand20203d}%
  \BibitemOpen
  \bibfield  {author} {\bibinfo {author} {\bibfnamefont {N.}~\bibnamefont
  {Harmand}}\ and\ \bibinfo {author} {\bibfnamefont {S.}~\bibnamefont
  {H{\'e}non}},\ }\href@noop {} {\bibfield  {journal} {\bibinfo  {journal}
  {arXiv preprint arXiv:2005.07589}\ } (\bibinfo {year} {2020})}\BibitemShut
  {NoStop}%
\bibitem [{\citenamefont {Turiv}\ \emph {et~al.}(2020)\citenamefont {Turiv},
  \citenamefont {Krieger}, \citenamefont {Babakhanova}, \citenamefont {Yu},
  \citenamefont {Shiyanovskii}, \citenamefont {Wei}, \citenamefont {Kim},\ and\
  \citenamefont {Lavrentovich}}]{turiv2020topology}%
  \BibitemOpen
  \bibfield  {author} {\bibinfo {author} {\bibfnamefont {T.}~\bibnamefont
  {Turiv}}, \bibinfo {author} {\bibfnamefont {J.}~\bibnamefont {Krieger}},
  \bibinfo {author} {\bibfnamefont {G.}~\bibnamefont {Babakhanova}}, \bibinfo
  {author} {\bibfnamefont {H.}~\bibnamefont {Yu}}, \bibinfo {author}
  {\bibfnamefont {S.~V.}\ \bibnamefont {Shiyanovskii}}, \bibinfo {author}
  {\bibfnamefont {Q.-H.}\ \bibnamefont {Wei}}, \bibinfo {author} {\bibfnamefont
  {M.-H.}\ \bibnamefont {Kim}}, \ and\ \bibinfo {author} {\bibfnamefont
  {O.~D.}\ \bibnamefont {Lavrentovich}},\ }\href@noop {} {\bibfield  {journal}
  {\bibinfo  {journal} {Science Advances}\ }\textbf {\bibinfo {volume} {6}},\
  \bibinfo {pages} {eaaz6485} (\bibinfo {year} {2020})}\BibitemShut {NoStop}%
\bibitem [{\citenamefont {Seung}\ and\ \citenamefont
  {Nelson}(1988)}]{seung1988defects}%
  \BibitemOpen
  \bibfield  {author} {\bibinfo {author} {\bibfnamefont {H.~S.}\ \bibnamefont
  {Seung}}\ and\ \bibinfo {author} {\bibfnamefont {D.~R.}\ \bibnamefont
  {Nelson}},\ }\href@noop {} {\bibfield  {journal} {\bibinfo  {journal}
  {Physical Review A}\ }\textbf {\bibinfo {volume} {38}},\ \bibinfo {pages}
  {1005} (\bibinfo {year} {1988})}\BibitemShut {NoStop}%
\end{thebibliography}%

\end{document}